\documentclass[aip,jcp,reprint,twocolumn,showpacs,amsmath,amssymb]{revtex4-1}

\usepackage[english]{babel}
\usepackage{graphicx}
\usepackage{float}
\usepackage{bm}
\usepackage[utf8]{inputenc}

\begin{document}

\title{Nuclear quantum effects in liquid water from path-integral simulations using an \textit{ab initio} force matching approach}
\author{Thomas Spura}
\author{Christopher John}
\address{Institute for Physical Chemistry, University of Mainz, Staudinger Weg 9, D-55128 Mainz, Germany}
\author{Scott Habershon}
\address{Department of Chemistry and Centre for Scientific Computing, University of Warwick, Coventry CV4 7AL, United Kingdom}
\author{Thomas D. K\"{u}hne}
\email{kuehne@uni-mainz.de}
\address{Institute for Physical Chemistry and Institute for Computational Sciences, University of Mainz, Staudinger Weg 7, D-55128 Mainz, Germany}
\date{\today}
%

%
%
\begin{abstract}
We have applied path integral simulations, in combination with new \textit{ab initio} based water potentials, to investigate nuclear quantum effects in liquid water. Because direct \textit{ab initio} path integral simulations are computationally expensive, a flexible water model is parameterized by force-matching to density functional theory-based molecular dynamics simulations. The resulting effective potentials provide an inexpensive replacement for direct \textit{ab inito} molecular dynamics simulations and allow efficient simulation of nuclear quantum effects. Static and dynamic properties of liquid water at ambient conditions are presented and the role of nuclear quantum effects, exchange-correlation functionals and dispersion corrections are discussed in regards to reproducing the experimental properties of liquid water.
\end{abstract}
\maketitle

%
%
\section{Introduction}

Liquid water is arguably one of the most important liquids due to its role in chemistry, biology and geophysics and, as such, also one of the most studied systems.\cite{RahmanStillinger1971} Despite this, a detailed understanding of the physical chemistry of water is still lacking due to its complex behaviour and unusual properties.\cite{Kauzmann1969} However, computational studies of water are rather challenging due to the presence of the various physical phenomena that conspire to make water unique, such as the cooperativity of the hydrogen bond (HB) network, large polarizability effects, strong permanent dipoles and sizeable nuclear quantum effects (NQE). \cite{StillingerScience}

The role of zero-point energy (ZPE) and tunnelling effects in modifying the strength of interactions in the HB network of ambient liquid water, and the consequences for the static and dynamic properties, has been appreciated for almost three decades now.\cite{kuharski1085studystructure,wallqvist1985pimd, miller2005complexmolecularsystems} Although it is well known that NQE generally weaken intermolecular hydrogen-bonding, resulting in a less-structured liquid and concomitantly faster rotational and translational dynamics, \cite{lobaugh1997quantumwater,paesani:184507,PaesaniVoth2009,miller:154504,hernandez2006modeldependence,comp_quant_eff} there is ongoing debate regarding the magnitude of this effect. For example, while comparisons of classical and quantum (path integral) simulations of liquid water using empirical force-fields generally predict that the rates of dynamic processes are increased by around 50\% due to NQE,\cite{lobaugh1997quantumwater,paesani:184507,miller:154504} recent simulations using a water model specifically parameterized for quantum simulations suggests an enhancement of just 15\%. \cite{comp_quant_eff}

An \textit{ab initio} PIMD approach, where the interatomic forces are calculated on-the-fly from accurate electronic structure calculations, would be very attractive to address the questions surrounding the role of NQE in liquid water. Considerable effort has gone into devising practical density functional theory (DFT) based PIMD methods \cite{marx:4077, tuckerman:5579} and much progress has been reported.\cite{PhysRevLett.91.215503, PhysRevLett.101.017801} Nevertheless, the computational expense of this route still severely limits the length- and time-scales that can be studied. 

In this work, we take a different route. To circumvent the computational cost associated with an \textit{ab initio} PIMD technique, we instead develop here a flexible water models which are derived by matching the interatomic forces to those from accurate electronic structure calculations,\cite{ercolessi2007force_matching} without relying on any empirical parameters or experimental input. This not only facilitates large-scale PIMD simulations with an accuracy that is similar to DFT-based PIMD calculations, but at variance to empirical force-fields that are parameterized to reproduce experimental data, is also not plagued by a``double-counting" of NQE.\cite{comp_quant_eff} Furthermore, this allows us to assess the accuracy and intrinsic properties of potential DFT-based PIMD simulations as distinct from those that arise from numerical approximations, insufficient sampling and finite-size effects. However, contrary to explicit electronic structure-based PIMD simulations, the employed functional form of the recently devised q-TIP4P/F force-field entails that the resulting water model is neither polarizable nor able to simulate chemical reactions that may take place in water.\cite{comp_quant_eff} 

The remainder of this paper is organized as follows. In Section~II, we present the force-matching scheme used to derive the parameters of new flexible q-TIP4P/F-like water models. The finite temperature path-integral methods used to rigorously account for ZPE and tunnelling effects, and to investigate the influence of NQE in liquid water are described in Section~III. Thereafter, in Section~IV, we describe computational details, and in Section~V we assess the accuracy of water models derived using the force-matching procedure. The eventual performance of our newly derived water models and the influence of NQE are discussed in Section~V, which is followed by conclusions in Section~VI.

%
%
\section{Force-Matching\label{sec:theory}}

Empirical water force-fields are typically parameterized so as to reproduce experimental data such as the radial distribution function (RDF), structure factor, heat of vaporization and the density maximum of liquid water.\cite{RahmanStillinger1974, jorgensen_TIP, spce, TIP5P, TIP4P2005, Guillot2002}
While these potentials are usually remarkably accurate in reproducing the underlying experiments, the transferability to regions of the phase diagram or situations different from that in which they have been fitted may be restricted. Furthermore, since NQE are already present in experiment, they will be considered twofold when taken explicitly into account within a PIMD simulation.\cite{miller2005quantumdiffusion} 

Using results from accurate \textit{ab initio} electronic structure calculations where, contrary to experimental data, NQE are not present, this ``double-counting'' of NQE is circumvented from the outset and permits to study the impact of NQE in a direct and systematic manner. Beside the finite-difference approach \cite{VegaAbascal2011} there are many schemes to fit empirical models to \textit{ab initio} data, including the inverse Monte Carlo \cite{IMC1,IMC2} or iterative Boltzmann inversion \cite{IBI} technique that both rely on Henderson's theorem, which states that a potential with only pairwise interactions is uniquely determined by the RDF up to an additive constant.\cite{Henderson1974} 
However, the application of Henderson's theorem is not without problems since at finite numerical accuracy essentially indistinguishable RDFs may entail very different pair potentials.\cite{Potestio2013} Furthermore, the generation of reference RDFs from first-principles by \textit{ab initio} MD (AIMD) simulations is computationally rather time consuming, \cite{laasonen1993ailiqwater, SprikWater1996, PhysRevE.68.041505, GrossmanWater2004, KuoWater2004, FernandezWater2004, JoostWater2005, SitMarzariWater2005, lee:154507, TodorovaWater2006, SchiffmannWater2008, schmidtNPT2009, kuehnewater2009, GuidonWater2010, BanyaiNMR2010, GalliSpanuWater2011, FernandezSerra2011, GalliVdWWater2011, chun2012structure, kuehne2ptwater2012, TuckermanWater2012, kuehnewater2013, WaterPNAS2013, kuehnewaterreview2013, MP2Water2013} in particular when considering many state points to guarantee that the resulting water model is as transferable as possible. 

In contrast, the force-matching technique of Ercolessi and Adams,\cite{ercolessi2007force_matching} where the interaction potential is derived so as to mimic the forces of accurate reference calculations, not only includes many-body environmental effects, but also allows to employ a higher level of theory since fewer electronic structure calculations are required, in general. To determine the parameters of an empirical interaction potential given \textit{ab initio} force calculations for a set of configurations, we minimize the normalized $L_1$ force distance $\|\delta\bm{F}\|_{1}$, 
\begin{equation}
\label{eq:chi}
	\|\delta\bm{F}\|_{1} = \frac{1}{3} \left< \sum_{i=1}^{N} \sum_{\alpha \in (x,y,z)} \left[ \frac{|\bm{F}_{i,\alpha}^{\text{QM}} - \bm{F}_{i,\alpha}^{\text{FF}}|}{\sigma_i} \right] \right>
\text{,}
\end{equation}
where $N$ is the number of atoms and $\sigma_i$ for the standard deviation of the force distribution $\bm{F}_{i, \alpha}$ of atom $i$ in directions $\alpha \in (x,y,z)$, while $\left< \cdot \cdot \cdot \right>$ implies the ensemble average of selected configurations from a PIMD simulation. The quantum mechanical reference forces are denoted as $\bm{F}_{i,\alpha}^{\text{QM}}$, while $\bm{F}_{i,\alpha}^{\text{FF}}$ are the nuclear forces of the classical interaction potential, respectively. 

In any case, the minimization of Eq.~\ref{eq:chi} with respect to the parameters of $\bm{F}_{i,\alpha}^{\text{FF}}$ represents an ill-posed problem, in particular when including atomic partial charges in the optimization procedure. From this it follows that the optimization may not be stable under small variations of the corresponding parameters. This is reflected in an error landscape with many saddle points and flat areas, where the Hessian matrix is nearly singular, which leads to inaccuracies due to the limited precision of floating point arithmetic. 
As a consequence, an important problem of gradient-based minimization methods is the particular form of the objective function, whose derivative with respect to partial charges are often found to be ill-conditioned. 

Even though it is possible to mitigate this difficulty by augmenting the penalty function with additional properties such as the total force or torque with its respective weights,\cite{akin-ojo2008qualityforcefields,sala2011fm, sala2012fm} here we propose to circumvent this using the sequential least-squares quadratic programming algorithm (SLSQP) together with physically-sensible bound constraints \cite{SLSQP}. The SLSQP method treats the original problem as a sequence of constrained least-squares problems that is equivalent to a quadratic programming algorithm for nonlinearly-constrained gradient-based optimization, hence the name. Specifically, each SLSQP step involves solving a quadratic approximation of the original objective function, where the linear term is the gradient and the quadratic term is an approximate Hessian, with first-order affine approximations of the nonlinear constraints. The approximate Hessian, which is initialized to the identity matrix, is continuously updated, while keeping it positive definite, based on the gradients and function values at subsequent steps similar to the BFGS quasi-Newton scheme.\cite{NumericalRecipes} As a consequence, like any quasi-Newton method, the true Hessian is only approached in the limit of many iterations close to the minimum. As a result of the ill-posed nature of the problem, we search for the minimum along the direction of the modified quasi-Newton scheme by first bracketing the minimum and then using Brent's method.\cite{Brent1973} At variance to more elaborate techniques that exploit gradient information, here the availability of the function's derivative is not required. However, it should be noted that this procedure offers no guarantees about whether the global minimum of the optimization function is located.

%
%
\subsection{Water model\label{params_desc}}

The aim of this work is to use the force matching procedure outlined above to fit simple empirical force-fields to \textit{ab initio} force data; this requires us to choose a functional form for the empirical water model, within which the parameters will be optimised. Among the large number of simple point charge models that have been developed for liquid water, we have chosen the flexible q-TIP4P/F water model of Habershon \textit{et al.} \cite{comp_quant_eff}, which has been shown to offer a good reproduction of several key experimental properties of liquid water under ambient conditions, including diffusion coefficients, liquid density and liquid structure.

The q-TIP4P/F water model consists of two positive charge sites of magnitude $|\frac{q}{2}|$ on the
hydrogen atoms and a negative charge of magnitude $q$ positioned at $\bm{r}_M = \gamma \bm{r}_O + (1-\gamma)(\bm{r}_{H_1}-\bm{r}_{H_2})/2$ to ensure local charge neutrality of each water molecule. These so-called M-sites and the hydrogen atoms on different water molecules interact with each other through a simple Coulomb potential. In conjunction with a Lennard-Jones potential between the oxygen atoms, this constitutes the following pairwise-additive intermolecular potential
\begin{eqnarray}
\nonumber
    V_{\text{inter}} &=& \sum_i \sum_{j>i} 4 \epsilon
    \left[\left(\frac{\sigma}{r_{ij}}\right)^{12} -
    \left(\frac{\sigma}{r_{ij}}\right)^6\right]
    \\
    &+& \sum_{m \in i} \sum_{n \in j}
                \frac{q_m q_n}{r_{mn}} 
    \text{,}
\end{eqnarray}
where $r_{ij}$ is the distance between the oxygen atoms and $r_{mn}$ the distance between the partial charges in molecules $i$ and $j$.

Flexibility is added to this model by an intramolecular potential, which consists of an anharmonic quartic expansion of the Morse potential and a harmonic bending term, 
\begin{eqnarray}
V_{\text{intra}} = \sum_i \left[ \frac{1}{2} k_{\theta} (\theta_i - \theta_{\text{eq}})^2 + V_{\text{OH}}(r_{i1}) + V_{\text{OH}}(r_{i2}) \right] \!\! , \, \quad
\end{eqnarray}
where
\begin{eqnarray}
V_{\text{OH}}(r) &=& D_r \Bigl[ \alpha_r^2 (r - r_{\text{eq}})^2 - \alpha_r^3 (r - r_{\text{eq}})^3 \nonumber \\
	      &+& \frac{7}{12} \alpha_r^4 (r - r_{\text{eq}})^4 \Bigr]. \nonumber
\end{eqnarray}
Here $r_{\text{eq}}$ denotes the intramolecular O-H equilibrium distance, $r_{i1}$ and $r_{i2}$ are the two covalent O-H bonds of water molecule $i$, $\theta_{\text{eq}}$ is the equilibrium H-O-H bond angle and $\theta_{i}$ is the H-O-H bond angle in molecule $i$.

In this work, the central aim is to modify the nine independent parameters of the original q-TIP4P/F water model such that it reproduces the forces determined in \textit{ab initio} calculations. In particular, we optimise these parameters for a series of different DFT functionals, resulting in several different q-TIP4P/F-like water models.

%
%
\section{Path Integral Formalism\label{sec:pimd}}

\subsection{Path Integral Molecular Dynamics}

In the path integral molecular dynamics (PIMD) method, each quantum particle is replaced by a classical harmonic $p$-bead ring-polymer. This extended system is isomorphic to the original quantum system, enabling calculation of quantum-mechanical properties of the system by sampling the path integral phase space. \cite{Feynman, chandler1981isomorphism, parrinello1984moltenkcl, CeperleyRMP} The canonical quantum partition function, $Z_{p}$, can be expressed in terms of the Hamiltonian $\hat H = \hat T + \hat V$ and the inverse temperature $\beta^{-1} = {k_B T}$, 
\begin{equation}
Z = \text{Tr}\left[ e^{-\beta \hat H} \right] = \text{Tr} \left[ \left( e^{-\beta_{p} \hat H} \right)^{p} \right] = \lim_{p \rightarrow
\infty}Z_p. \label{QuantPartFunc}
\end{equation} 
Inserting $p-1$ complete sets of position eigenstates, and using the symmetric Trotter splitting to represent the Boltzmann operator, Eq.~(\ref{QuantPartFunc}) can be written in a computationally convenient form, which can be directly sampled using the Monte Carlo technique, as 
\begin{eqnarray}
Z_p &=& \left(\frac{m}{2 \pi \beta_p }\right)^{\frac{3p}{2}}\int\!d^p \, \bm{r} \\ 
&\times& e^{-\beta_p \sum \limits_{k=1}^{p}
\big{[}\frac{1}{2}m\omega^2_p(\bm{r}^{(k)}-\bm{r}^{(k+1)})^2+V(\bm{r}^{(k)})\big{]}_{\bm{r}^{(p+1)}=\bm{r}^{(1)} }}
\text{,} 
\nonumber
\end{eqnarray}
where $p$ is the number of imaginary time slices, $m$ the particle mass and $\omega_p=p/\beta = 1/\beta_p$ the angular frequency of the harmonic spring potential between adjacent beads. The constraint $\bm{r}^{(p+1)} = \bm{r}^{(1)}$, where the parenthesis in the exponent denotes the bead index, is a result of the trace in Eq.~(\ref{QuantPartFunc}) and means that the corresponding $p$-bead system is a closed ring-polymer, while $\lim_{p \rightarrow \infty}{Z}_p={Z}$ is a direct consequence of the Trotter theorem, which states that 
\begin{equation}\label{trotter}
e^{\alpha (\hat A+ \hat B)} =
\lim_{p\rightarrow\infty}[e^{\frac{\alpha}{2p}\hat
B}e^{\frac{\alpha}{p}\hat A}e^{\frac{\alpha}{2p}\hat B}]^p \text{.}
\end{equation}
The latter implies that in the limit $p\rightarrow\infty$ the solution of sampling $Z_p$ classically is equivalent to the exact quantum partition function. \cite{parrinello1984moltenkcl}

Making use of the standard Gaussian integral to introduce momenta, $Z_p$ can be also be sampled using MD. If we further generalize the resulting expression for more than one particle, the quantum partition function eventually reads as
\begin{eqnarray}
Z_p&=&\mathcal{N}\int\!d^{Np}\,\bm{r} \int\! d^{Np}\,\bm{p} \; e^{- \beta_{p} H_p(\{\bm{r}\}, \{\bm{p}\})}, 
\end{eqnarray}
where $\mathcal{N}$ is a normalisation constant and 
\begin{eqnarray}
\nonumber
 H_p(\{\bm{r}\}, \{\bm{p}\}) &=& \sum_{k=1}^{p} \left[\sum_{i=1}^{N}\left(\frac{(\bm{p}_i^{(k)})^2}{2 m_i^{(k)'}}+\frac{m_i\omega_p^2}{2}(\bm{r}_i^{(k)} - \bm{r}_i^{(k+1)})^2\right)\right] \\ &+& V(\bm{r}_1^{(k)}, ...,\bm{r}_N^{(k)}) 
\end{eqnarray}
is the so-called bead-Hamiltonian that describes the interactions between all $N$ particles of a system and for all $p$ beads. Finally, we note that time-independent quantum thermal properties of position-dependent operators can now be calculated straightforwardly in PIMD simulations according to
\begin{eqnarray}
\langle A \rangle_p &=& \frac{\mathcal{N}}{Z_{p}}\int\!d^{Np}\,\bm{r} \int\! d^{Np}\,\bm{p} \; e^{- \beta_{p} H_p(\{\bm{r}\}, \{\bm{p}\})} A_{p}(\mathbf{r}), 
\end{eqnarray}
where $A_{p}(\mathbf{r})$ is given as the bead-average of the operator $\hat{A}$, thus
\begin{equation}
A_{p}(\mathbf{r}) = \frac{1}{p} \sum_{k=1}^{p} A( \mathbf{r}^{(k)}).
\end{equation}
By comparing classical ($p=1$) and PIMD simulations, this approach allows one to assess the impact of NQE in time-independent observables such as RDFs.

In order to reduce the computational effort required to calculate the long-range electrostatic interactions $p$ times, we use the ring-polymer contraction scheme of Markland and Manolopoulos.\cite{Markland2008256} Here, we split the Hamiltonian into its inter- and intramolecular contributions and limit the computationally-expensive intermolecular force calculation to a single Ewald sum at the centroid of the ring-polymer system:
\begin{equation}
 \bm{r}_i^{(c)}= \frac{1}{p}\sum_{k=1}^p\bm{r}_i^{(k)}\text{.}
\end{equation} 
Short-range corrections are subsequently added to account for the impact of this approximation on the actual ring-polymer beads.

%
%
\subsection{Ring Polymer Molecular Dynamics}

In contrast to the original PIMD approach, the ring-polymer MD (RPMD) scheme of Craig and Manolopoulos allows one to approximate dynamical properties within the path-integral framework\cite{craig:3368,RPMDreview}. The diffusion coefficient, for instance, is obtained as the time-integral of the Kubo-transformed velocity auto-correlation function $\tilde{c}_{vv} (t)$,
\begin{equation}\label{Diffusion}
D = \frac{1}{3} \int_0^{\infty} dt \, \tilde c_{vv} (t).
\end{equation}
The RPMD method approximates the quantum-mechanical Kubo-transformed time-correlation function $\tilde c_{AB} (t)$ as a classical time-correlation function calculated in the extended path integral phase-space. Thus, in RPMD, we have
\begin{eqnarray}
\label{RPMDcorrfunc}
\tilde{c}_{AB}(t) &\approx& \frac{\mathcal{N}}{Z_p}
\int\! d^{Np}\bm{p}\, d^{Np}\bm{r} \\
&\times& e^{-\beta_p H_p(\{\bm{r}\},\{\bm{p}\})} A_p(\{\bm{r}(0)\})B_p(\{\bm{r}(t)\})\, ,  
\nonumber
\end{eqnarray}
where
\begin{subequations}
\begin{eqnarray}
B_p(\{\bm{r}(t)\})&=&\frac{1}{p}\sum\limits_{k=1}^p
  B(\bm{r}_1^{(k)}(t),\ldots,\bm{r}_N^{(k)}(t)) 
\end{eqnarray}
and
\begin{eqnarray}
A_p(\{\bm{r}(0)\})&=&\frac{1}{p}\sum\limits_{k=1}^p
  A(\bm{r}_1^{(k)}(0),\ldots,\bm{r}_N^{(k)}(0))
\end{eqnarray}
\end{subequations}
are ensemble averages over the beads of a closed ring-polymer. Manolopoulos and coworkers have shown that this approximation is exact in the high-temperature limit, where Eq.~\ref{RPMDcorrfunc} reduces to the classical correlation function, and also in the short-time and simple harmonic oscillator limits \cite{craig:3368, habershon2009quantumleakage, RPMDreview}. In this work, RPMD is used to calculate molecular diffusion coefficients for each of the water models developed by our force-matching approach.

However, to circumvent the spurious vibrational modes which arise from the internal ring-polymer modes in RPMD simulations \cite{habershonIR}, simulations of vibrational spectra in this work employ the Partially Adiabatic Centroid Molecular Dynamic (PACMD) method. \cite{hone:154103}. In this approach, the effective masses of the ring-polymer beads are adjusted so as to shift the spurious oscillations beyond the spectral range of interest \cite{parrinello1984moltenkcl}. Specifically, the elements of the Parrinello-Rahman mass matrix are chosen so that the internal modes of the ring-polymer are shifted to a frequency of
\begin{equation}
\Omega = \frac{p^{p/p-1}}{\beta \hbar}, 
\end{equation} 
which allows for similar integration time-steps to be used in both RPMD and PACMD simulations.\cite{habershonIR}

%
%
\section{Computational Details}

In attempting to generate empirical water models that are as transferable as possible, we have extracted 1500 decorrelated snapshots from PIMD simulations consisting of 125 water molecules in the constant-NPT (isothermal,isobaric) ensemble using the q-TIP4P/F water potential of Habershon \textit{et al.} \cite{comp_quant_eff}. Specifically, we have selected 125 different configurations at 1~bar pressure for each temperature over the whole liquid temperature range between $248~\text{K}$ to $358~\text{K}$ in $10~\text{K}$ steps. In this way, the resulting water model is not just parametrized to a single state point at ambient conditions but spans a range of state points from undercooled water to near the vapor phase. 

Force matching, as described in Section~\ref{sec:theory}, was conducted based on reference forces from DFT calculations. We employed the mixed Gaussian and plane wave approach \cite{lippert1997gaussianplanewave} as implemented in the \textsc{CP2K/Quickstep} code \cite{VandeVondele_quickstep}. In this approach the Kohn-Sham orbitals are represented by a TZV2P Gaussian basis set \cite{vandevondele2007gaussianbasis}, while the charge density is expanded in plane waves using a density cutoff of 320~Ry. The exchange and correlation (XC) energy was described by a series of common generalized gradient approximations, and norm-conserving Goedecker-Teter-Hutter pseudopotentials were used to describe the interactions between the valence electrons and the ionic cores \cite{GTH1996pseudo,hartwigsen1998separabledualspacegaussian,KrackGTH}. Van der Waals (vdW) interactions, which are typically left out by common local and semi-local XC functionals, are either approximated by an additional pair-potential, or by dispersion-corrected atom-centered pseudopotentials (DCACP) \cite{grimme2010d3, lilienfeld2005dcacp}.

%
\begin{table*}
\caption{Parameters of q-TIP4P/F-like water models obtained with the force-matching approach.}\label{tab:parameters}
\begin{tabular}{l|c|c|c|c|c|c|c|c|c}
    XC Functional & $q$ [e] & $\gamma$ & $\sigma$ [$a_{0}$] & $\epsilon$ [$E_{h}$] & $ \theta_{\text{HOH}}$ [deg] & $r_{\text{OH}}$[$a_{0}$] & $D_r$ [$E_{h}$] & $k_\theta/2$ [$E_{h}$/deg$^2$] & $\alpha_{r}$ [1/$a_{0}$] \\
\hline
\hline
B97G \cite{B97, B97Grimme}  & -1.1437 & 0.65603 & 6.0330 & 2.1035$\times 10^{-4}$ & 107.42 & 1.8099 & 0.13773 & 6.2700$\times 10^{-2}$ & 1.3671 \\
B97G-D3    & -1.1228 & 0.65798 & 6.0122 & 2.2092$\times 10^{-4}$ & 107.42 & 1.8103 & 0.13670 & 6.2647$\times 10^{-2}$ & 1.3701 \\
BLYP \cite{becke, lyp}    & -1.0891 & 0.65468 & 6.0025 & 2.1267$\times 10^{-4}$ & 107.44 & 1.8296 & 0.13280 & 6.2954$\times 10^{-2}$ & 1.3509 \\
BLYP-D3     & -1.0738 & 0.65880 & 5.9702 & 2.3220$\times 10^{-4}$ & 107.40 & 1.8301 & 0.16625 & 6.2796$\times 10^{-2}$ & 1.2089 \\
BLYP-DCACP       & -1.0806 & 0.65194 & 5.9925 & 2.1913$\times 10^{-4}$ & 107.44 & 1.8320 & 0.13319 & 6.2799$\times 10^{-2}$ & 1.3432 \\
BP86 \cite{becke, perdew86}    & -1.1439 & 0.65123 & 5.9734 & 2.2413$\times 10^{-4}$ & 107.41 & 1.8297 & 0.16232 & 6.1953$\times 10^{-2}$ & 1.2282 \\
BP86-D3     & -1.1316 & 0.65539 & 5.9725 & 2.2810$\times 10^{-4}$ & 107.41 & 1.8297 & 0.16116 & 6.1854$\times 10^{-2}$ & 1.2320 \\
BP86-DCACP        & -1.1309 & 0.64901 & 5.9723 & 2.2765$\times 10^{-4}$ & 107.41 & 1.8330 & 0.15981 & 6.1897$\times 10^{-2}$ & 1.2308 \\
PBE \cite{pbe}     & -1.1347 & 0.65551 & 5.9746 & 2.2681$\times 10^{-4}$ & 107.41 & 1.8277 & 0.16249 & 6.1706$\times 10^{-2}$ & 1.2324 \\
PBE-D3      & -1.1309 & 0.65681 & 5.9745 & 2.2861$\times 10^{-4}$ & 107.41 & 1.8276 & 0.16199 & 6.1640$\times 10^{-2}$ & 1.2341 \\
PBE-DCACP         & -1.1357 & 0.65528 & 5.9732 & 2.2328$\times 10^{-4}$ & 107.41 & 1.8277 & 0.16274 & 6.1789$\times 10^{-2}$ & 1.2317 \\
revPBE \cite{revPBE}  & -1.1042 & 0.66934 & 6.0272 & 2.1281$\times 10^{-4}$ & 107.38 & 1.8223 & 0.13504 & 6.1977$\times 10^{-2}$ & 1.3600 \\
revPBE-D3   & -1.0992 & 0.67121 & 6.0258 & 2.1496$\times 10^{-4}$ & 107.37 & 1.8222 & 0.13414 & 6.1935$\times 10^{-2}$ & 1.3642 \\
revPBE-DCACP      & -1.1022 & 0.66937 & 6.0134 & 2.1512$\times 10^{-4}$ & 107.38 & 1.8226 & 0.13748 & 6.1923$\times 10^{-2}$ & 1.3472 \\
TPSS \cite{tpss}       & -1.0552 & 0.71151 & 6.0645 & 2.0568$\times 10^{-4}$ & 107.21 & 1.8265 & 0.12072 & 6.3820$\times 10^{-2}$ & 1.4260 \\
TPSS-D3     & -1.0318 & 0.72981 & 5.9782 & 2.5136$\times 10^{-4}$ & 107.38 & 1.8273 & 0.16199 & 6.3405$\times 10^{-2}$ & 1.2316 \\
\hline\hline
q-TIP4P/F \cite{comp_quant_eff}   & -1.1128 & 0.73612 & 5.9694 & 2.9515$\times 10^{-4}$ & 107.40 & 1.7800 & 0.185 & 7.0000$\times 10^{-2}$ & 1.2100 \\
\end{tabular}
\end{table*}

The parameters of the q-TIP4P/F-like water potentials were obtained by minimizing Eq.~\ref{eq:chi} using the SLSQP algorithm of Kraft with a convergence tolerance of $10^{-6}$ on the penalty function between different iterations \cite{SLSQP}. Gradients with respect to the various optimization parameters were computed using finite differences with a displacement of $10^{-8}$. The initial parameters were taken from the original q-TIP4P/F water model \cite{comp_quant_eff}, while the optimized parameters for the various XC functionals we have considered here are listed in Tab.~\ref{tab:parameters}.

The resulting water models are denoted as ``fm-TIP4P/F-XC'', where ``XC'' represents the employed XC functional of the DFT-based reference calculations. Unless stated otherwise, all of our PIMD calculations were performed at a temperature of 298~K and a pressure of 1~bar in the constant-NPT ensemble using 125 water molecules in a cubic simulation box. Periodic boundary conditions were applied using the minimum image convention. Short-range interactions were truncated at 9~\AA \, and Ewald summation was employed to calculate the long-range electrostatic interactions. The ring-polymer contraction scheme with a cut-off value of $\sigma$ = 5~\AA  \, was employed to reduce the electrostatic potential energy and force evaluations to single Ewald sum, thereby significantly speeding up the calculations \cite{Markland2008256}. Specifically, $p$ = 32~beads were employed, while the computationally expensive part of the electrostatic interactions were contracted to the centroid, which in the following is indicated as $p = 32 \to 1$. The evolution of the ring-polymer in time was performed analytically in the normal mode representation by a multiple time-step algorithm using a discretized time-step of 1.0~fs for the intermolecular and 0.125~fs for the intramolecular forces \cite{RESPA}. For comparison, additional simulations with classical nuclei were also performed ($p$=1). 
In all simulations, the system was pre-equilibrated in the constant-NVT ensemble for 50~ps followed by a 100~ps equilibration in the constant-NPT ensemble using an Andersen thermo- and an anisotropic Berendsen barostat, respectively  \cite{AndersenThermo, BerendsenBaro}. Ensemble averages were then computed over an additional $5~\text{ns}$ PIMD trajectory.

Two-phase simulations were performed to calculate the melting point of water \cite{BonevGalli2004}. For this purpose, direct coexistence simulations of the water-ice interface were performed under atmospheric pressure \cite{BrykHaymet2002, fernandez:144506}. The initial hexagonal ice configurations were generated by placing the oxygen atoms at their crystallographic sites \cite{HaywardReimers1997}, and determining the positions of the hydrogen atoms using the Monte Carlo procedure of Buch \textit{et al.} \cite{BuchIceMC1998} in such a way that the Bernal-Fowler rules \cite{BernalFowler1933, Petrenko1999} were satisfied and the total dipole moment of the simulation cell was exactly zero. 
The initial 288 molecules ice configuration was equilibrated in the presence of an Andersen thermostat and an anisotropic Berendsen barostat for $50~\text{ps}$ before putting the secondary prismatic ($1\bar{2}10$) face of the ice cell in direct contact with a separately equilibrated water system consisting of 280 molecules \cite{Furukawa2005}. 
Finally, the combined solid/liquid system consisted of 568 water molecules and was simulated for $10~\text{ns}$. 

The velocity autocorrelation function $\tilde c_{vv} (t)$ in Eq.~\ref{Diffusion} was calculated for 5~ps by time averaging over 100 consecutive constant-NVE RPMD trajectories of length 10~ps. After an initial equilibration in the constant-NVT ensemble for 100~ps, the momenta were resampled between each constant-NVE RPMD trajectory and the system re-equilibrated for another 2~ps before correlation functions were accumulated. 

Infrared (IR) spectra were calculated using the PACMD method by averaging over 300 constant-NVE PACMD trajectories, each of 20~ps length. Here, a time-step of 0.5~fs for the intermolecular and 0.1~fs for the intramolecular interactions was employed. After an initial equilibration in the constant-NVT ensemble for 100~ps, the momenta were resampled and the system re-equilibrated for another 2~ps between each constant-NVE PACMD trajectory. 

To assess the accuracy of our force matching procedure, an explicit 50~ps long classical ($p$=1) AIMD simulation was performed using the second-generation Car-Parrinello scheme of K\"uhne \textit{et al.} \cite{CP2G, CP2Greview}. The nuclear forces were computed at the DFT level using the PBE XC functional and otherwise the exactly same settings as before. This calculation, denoted as ``125 Water (PBE)'', was conducted in the constant-NVT ensemble at $300~\text{K}$ employing the thermostat of Bussi \textit{et al.} \cite{bussi2007csvr} with a time constant of $25.0~\text{fs}$.

%
%
\section{Assessment of force-matched water potentials \label{sec:application}}

Before studying the static and dynamic properties of the force-matched water models derived here, it is worth considering the optimised parameters, as shown in Table 1. We see that, while the $M$-site charge parameter $q$ tends to be similar to that of the original q-TIP4P/F model, the parameter determining the position of the $M$-site, namely $\gamma$, is in general smaller than that of q-TIP4P/F; as a result, we expect that the average dipole moments of the water molecules in the force-matched potentials will be slightly smaller than in q-TIP4P/F water. However, we note that decreasing $\gamma$ has the effect of increasing the tetrahedral quadrupole moment of the water molecules, and hence may promote tetrahedral structuring; this is consistent with the fact that the DFT-based water simulations, which were used as force input in this work, tend to be over-structured. Another interesting trend is seen in the Lennard-Jones parameter $\epsilon$, which is generally smaller than that found in q-TIP4P/F; this most likely arises to balance the increased structure caused by the increased tetrahedral quadrupole moments of the force-matched potentials, as noted above. Finally, we see that the intramolecular potential parameters in the new force-matched models suggest that the intramolecular modes may be slightly ``softer'' than q-TIP4P/F; the difference here must arise from the differing parameterisation approaches adopted for the different models, and possibly reflects the fact that the new water models were derived by force-matching to sampled water configurations while q-TIP4P/F was not.

%
\begin{figure}
    \centering
    \includegraphics[width=0.5\textwidth]{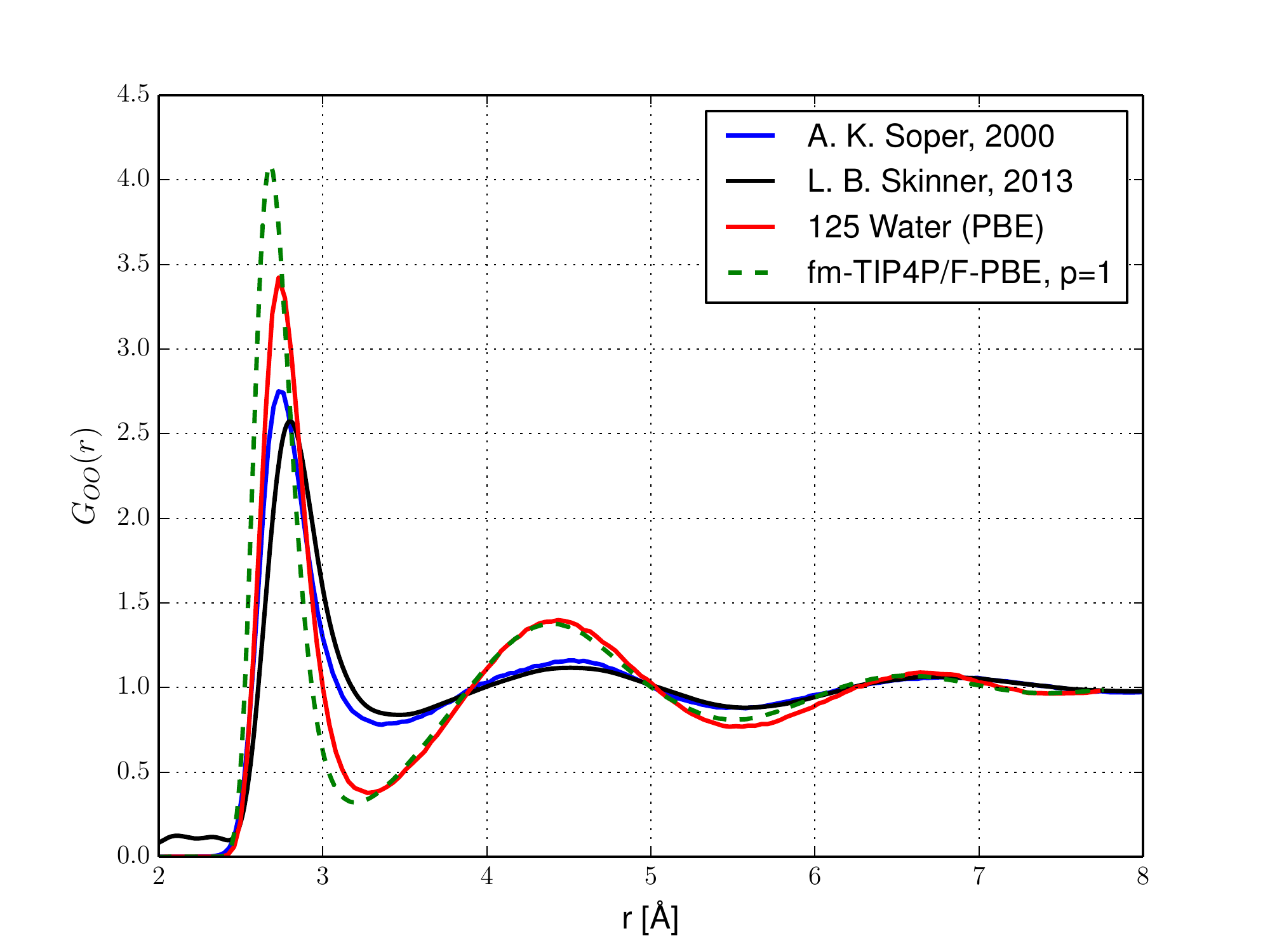}
    \caption{Oxygen-Oxygen of the fm-TIP4P/F-PBE water model and a DFT-based AIMD simulation. The experimental RDFs from Refs.~\onlinecite{soper2000radial} and \onlinecite{skinner2013oo} are shown for comparison. \label{fig:ref_PBE_OO}}
\end{figure}

To assess the quality of our force-matching procedure, we began by comparing the partial RDFs,\cite{kuehnegofr2013} as obtained by a classical MD simulation (p=1) using the fm-TIP4P/F-PBE potential with the corresponding DFT-based AIMD reference. The resulting O-O RDF are shown in Fig.~\ref{fig:ref_PBE_OO} and compared with recent neutron and x-ray diffraction measurements.\cite{soper2000radial,skinner2013oo} As can be seen the comparison with the experimental data reveals the well known overstructuring of DFT-based AIMD simulations.\cite{schmidtNPT2009, kuehnewater2009, BanyaiNMR2010, FernandezSerra2011, chun2012structure, kuehne2ptwater2012, TuckermanWater2012, kuehnewater2013, WaterPNAS2013, kuehnewaterreview2013} However, it also shows that the fm-TIP4P/F-PBE water model slightly underestimates the average O-O bond length and overestimates the height of the first peak within the O-O RDF with respect to the AIMD reference, whereas the second solvation shells are in excellent quantitative agreement. The O-H and H-H RDF, respectively, are shown as Figs.~S1 and S2 in the supporting information. The remaining error in the short-range portion of the RDFs are clearly most likely due to the simplicity of the force-matched potential, notably the exclusion of explicit polarisability, which is captured in the DFT simulations. Nevertheless, these results are promising, particularly considering that van der Waals interactions \cite{schmidtNPT2009, FernandezSerra2011, chun2012structure, TuckermanWater2012, kuehnewaterinterface2011} and  inclusion of NQE \cite{lobaugh1997quantumwater,paesani:184507,PaesaniVoth2009,miller:154504,hernandez2006modeldependence,comp_quant_eff, PhysRevLett.91.215503, PhysRevLett.101.017801} would be expected to improve agreement with experiment.

%
%
\subsection{Impact of Nuclear Quantum Effects}

%
\begin{figure}
    \centering
    \includegraphics[width=0.5\textwidth]{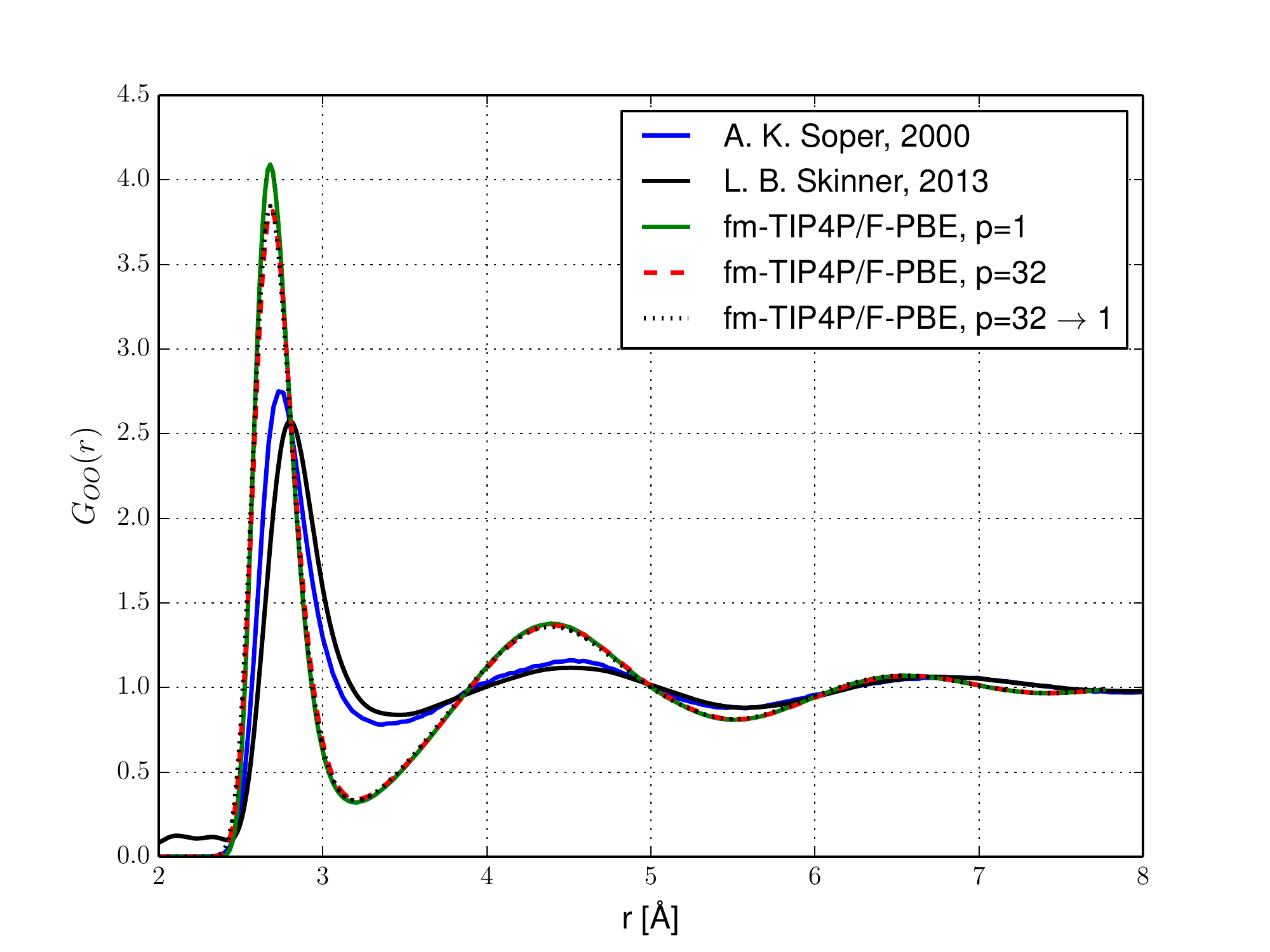}
    \caption{Oxygen-Oxygen RDF from classical MD and PIMD simulations using the fm-TIP4P/F-PBE water model. The experimental RDFs from Refs.~\onlinecite{soper2000radial} and \onlinecite{skinner2013oo} are shown for comparison\label{fig:nb_PBE_OO}.}
\end{figure}

%
\begin{figure}
    \centering
    \includegraphics[width=0.5\textwidth]{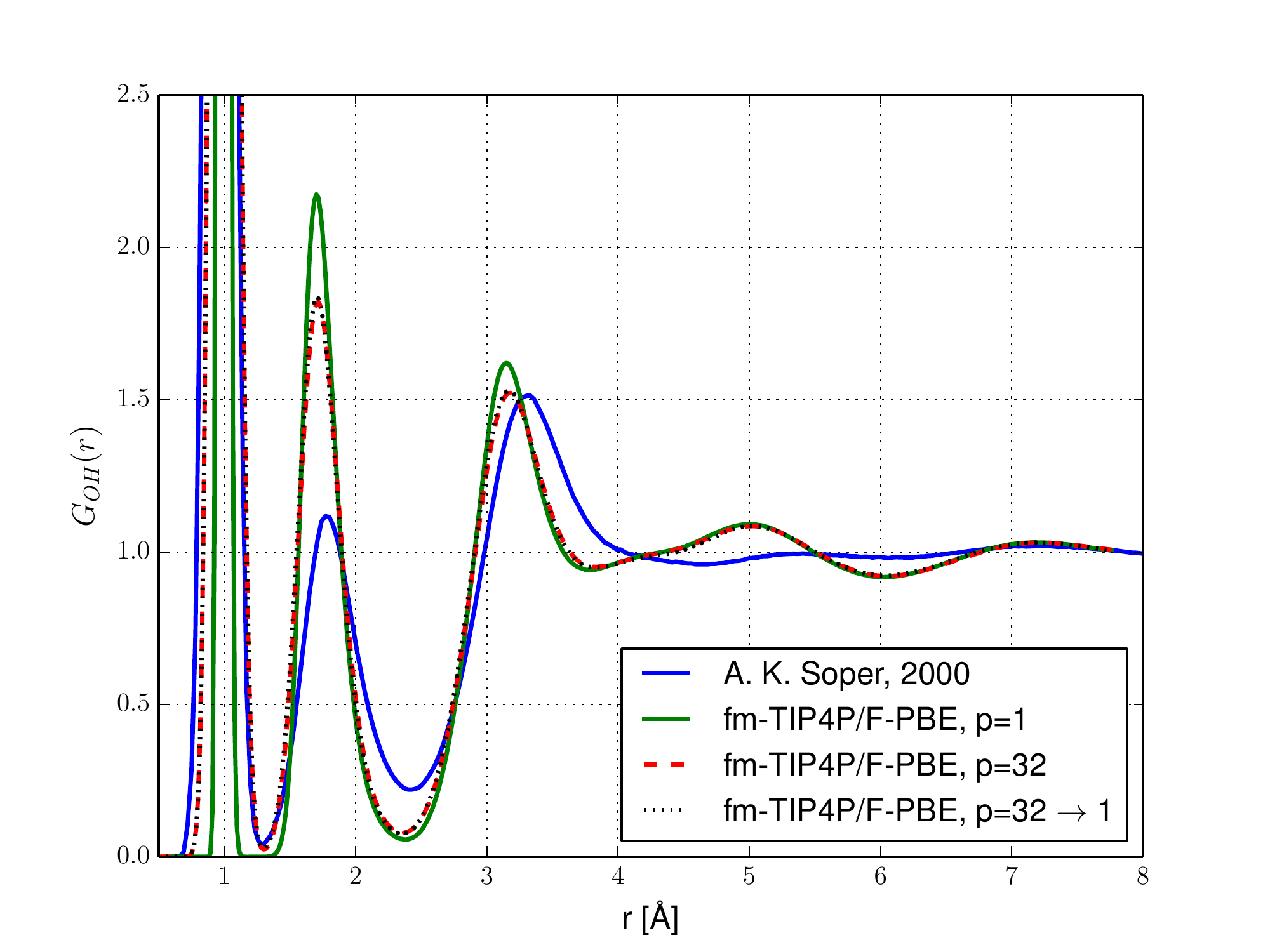}
    \caption{Oxygen-Hydrogen RDF from classical MD and PIMD simulations using the fm-TIP4P/F-PBE water model. The experimental RDFs from Ref.~\onlinecite{soper2000radial} is shown for comparison\label{fig:nb_PBE_OH}.}
\end{figure}

%
\begin{figure}
    \centering
    \includegraphics[width=0.5\textwidth]{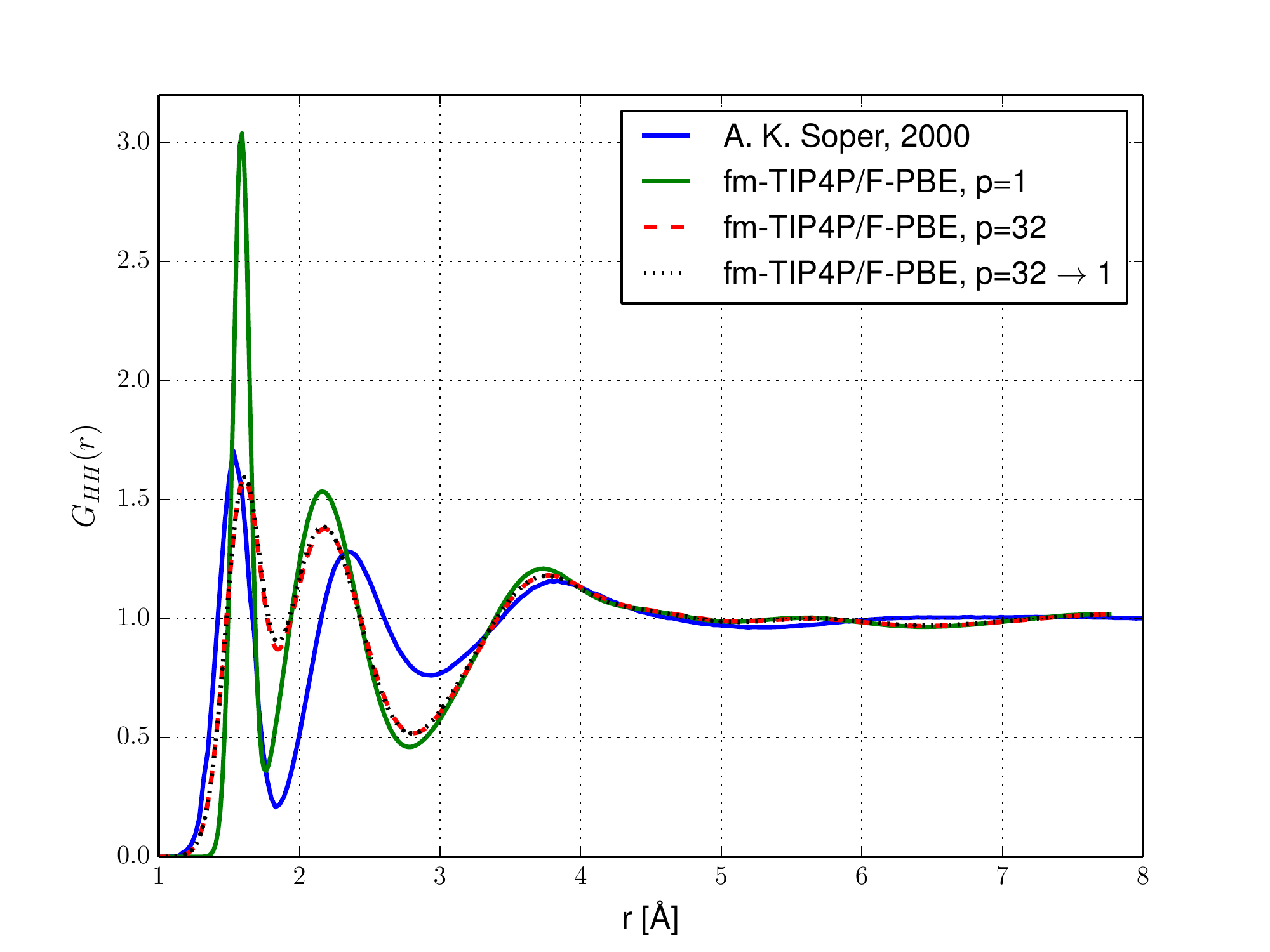}
    \caption{Hydrogen-Hydrogen RDF from classical MD and PIMD simulations using the fm-TIP4P/F-PBE water model. The experimental RDFs from Ref.~\onlinecite{soper2000radial} is shown for comparison\label{fig:nb_PBE_HH}.}
\end{figure}

To investigate the impact of NQE on the structure of liquid water, and to assess the approximation due to the ring-polymer contraction scheme in our force-matched models, we employed PIMD simulations. The corresponding results are displayed in Figs.~\ref{fig:nb_PBE_OO}, \ref{fig:nb_PBE_OH} and \ref{fig:nb_PBE_HH}, respectively. As expected, the inclusion of NQE softens the liquid water structure and, for the fm-TIP4P/F-PBE model, improves the agreement simulated and experimental RDFs. While the importance of NQE on the O-O RDF is rather small, they are clearly much more important whenever light atoms such as hydrogen are involved. The latter is a direct consequence of the fact that the radius-of-gyration of the (free) ring-polymer, which is a measure for the delocalization of the nuclear wave function, scales as $1/\sqrt{MT}$, where $M$ is the atomic mass and $T$ the nuclear temperature, and as such a clear manifestation that even at room temperature liquid water is a mild quantum fluid. The implications are particularly apparent in Fig.~\ref{fig:nb_PBE_OH}, where the delocalization of the average intramolecular O-H bond length is substantially increased, in agreement with experiment, as well as in Fig.~\ref{fig:nb_PBE_HH} where the height of the first peak is significantly reduced by quantum delocalisation. However, NQE had only a minor effect on the average bond lengths, so that all bonds are still somewhat too short compared to experiment. Finally, it is evident that the results using the ring-polymer contraction scheme ($p=32 \rightarrow 1$) are almost indistinguishable from explicit PIMD simulations ($p=32$), and is thus exclusively employed in the following.

%
%
\subsection{Influence of van der Waals interactions}

%
\begin{figure}
    \centering
    \includegraphics[width=0.5\textwidth]{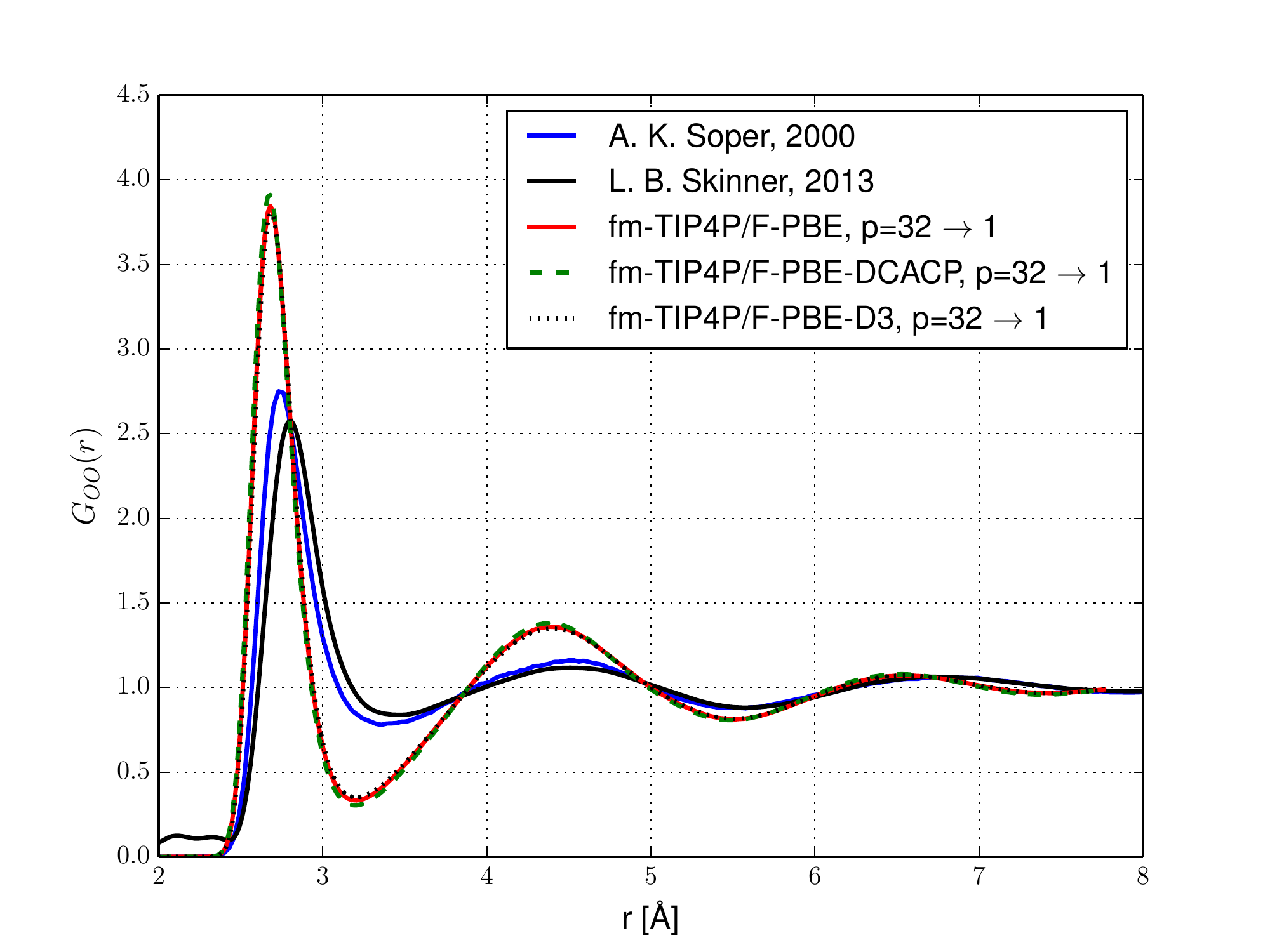}
    \caption{Oxygen-Oxygen RDF from PIMD simulations using the fm-TIP4P/F-PBE water model with and without London dispersion corrections. The experimental RDF from Refs.~\onlinecite{soper2000radial} and \onlinecite{skinner2013oo} are shown for comparison\label{fig:vdw_PBE}.}
\end{figure}

Since long-range vdW interactions are typically neglected by common local and semi-local XC functionals, we investigated to what extent approximate London dispersion correction schemes to DFT, such as DCACP and the ``D3'' correction of Grimme and coworkers, could improve the structure of liquid water. \cite{lilienfeld2005dcacp, grimme2010d3} The corresponding O-O RDFs are shown in Fig.~\ref{fig:vdw_PBE}, while the O-H and H-H are displayed in the supporting information as Fig.~S3 and S4, respectively. It is apparent that with the inclusion of NQE, both vdW correction schemes exhibit a marginal improvement in the RDFs. Nevertheless, due to the fact that both schemes systematically improve the agreement with experiment, from now on only results including the ``D3'' vdW correction will be presented, in particular since the latter have been shown to also improve the density and the translational diffusion of liquid water. \cite{schmidtNPT2009, kuehnewaterinterface2011, TuckermanWater2012, chun2012structure}

%
%
\subsection{Effect of the exchange-correlation functional}

%
\begin{figure}
    \centering
    \includegraphics[width=0.5\textwidth]{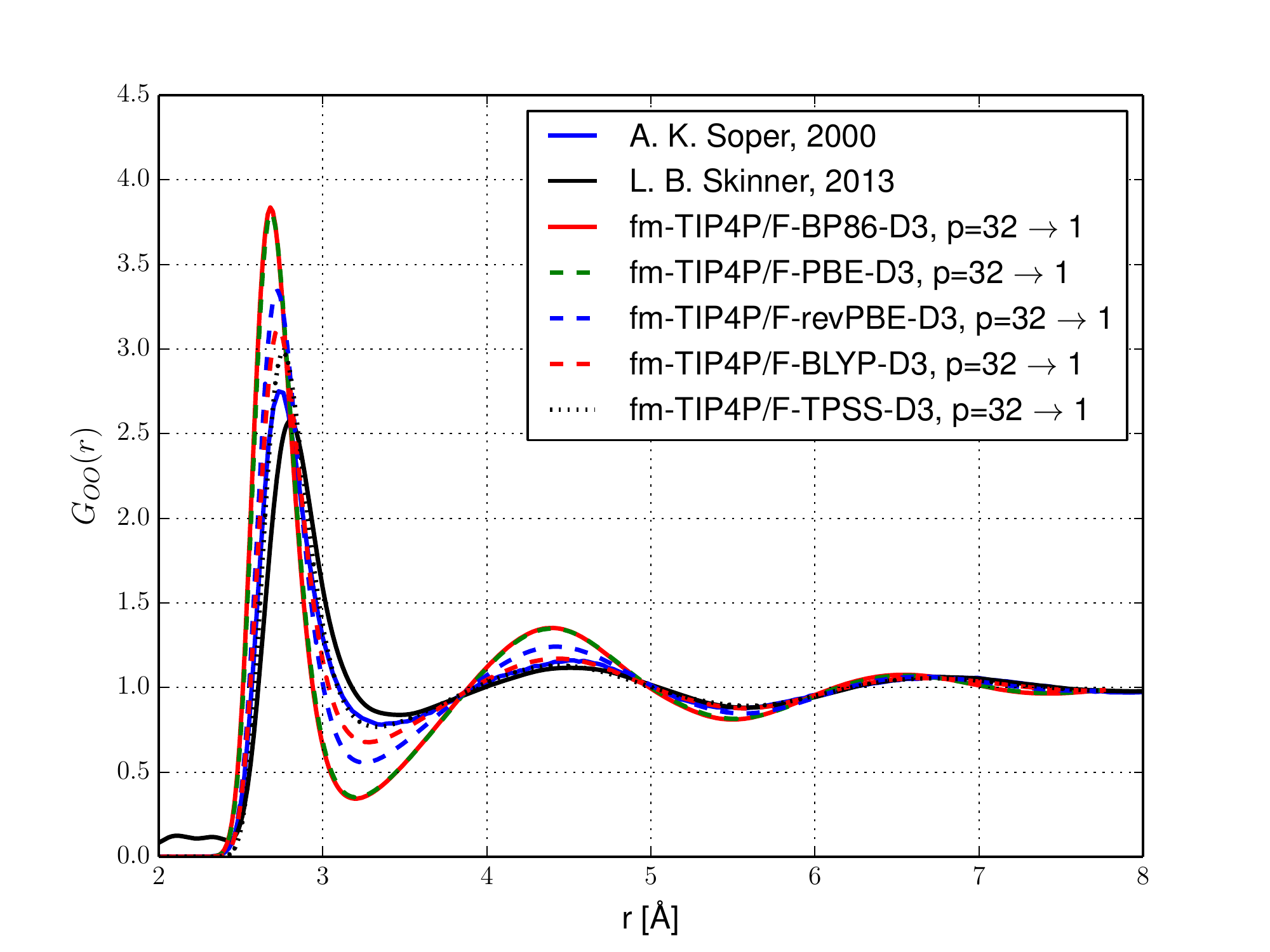}
    \caption{Oxygen-Oxygen RDFs from PIMD simulations using the fm-TIP4P/F-XC-D3 water model for the BP86, BLYP, revPBE, PBE and TPSS XC functionals, respectively. The experimental RDFs from Refs.~\onlinecite{soper2000radial} and \onlinecite{skinner2013oo} are shown for comparison\label{fig:comp_diff_func_OO}.}
\end{figure}

%
\begin{figure}
    \centering
    \includegraphics[width=0.5\textwidth]{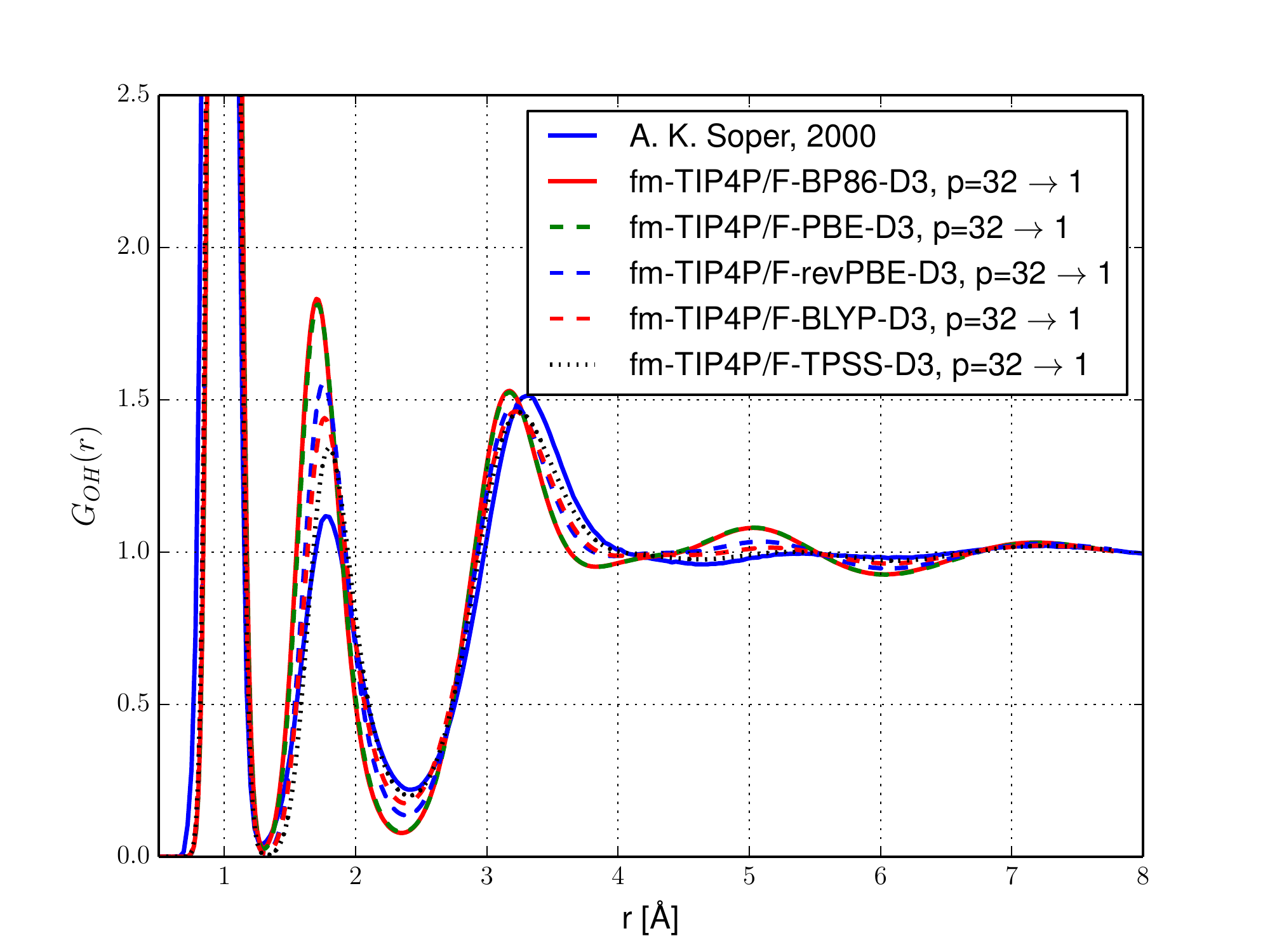}
    \caption{Oxygen-Hydrogen RDFs from PIMD simulations using the fm-TIP4P/F-XC-D3 water model for the BP86, BLYP, revPBE, PBE and TPSS XC functionals, respectively. The experimental RDFs from Ref.~\onlinecite{soper2000radial} is shown for comparison\label{fig:comp_diff_func_OH}.}
\end{figure}

%
\begin{figure}
    \centering
    \includegraphics[width=0.5\textwidth]{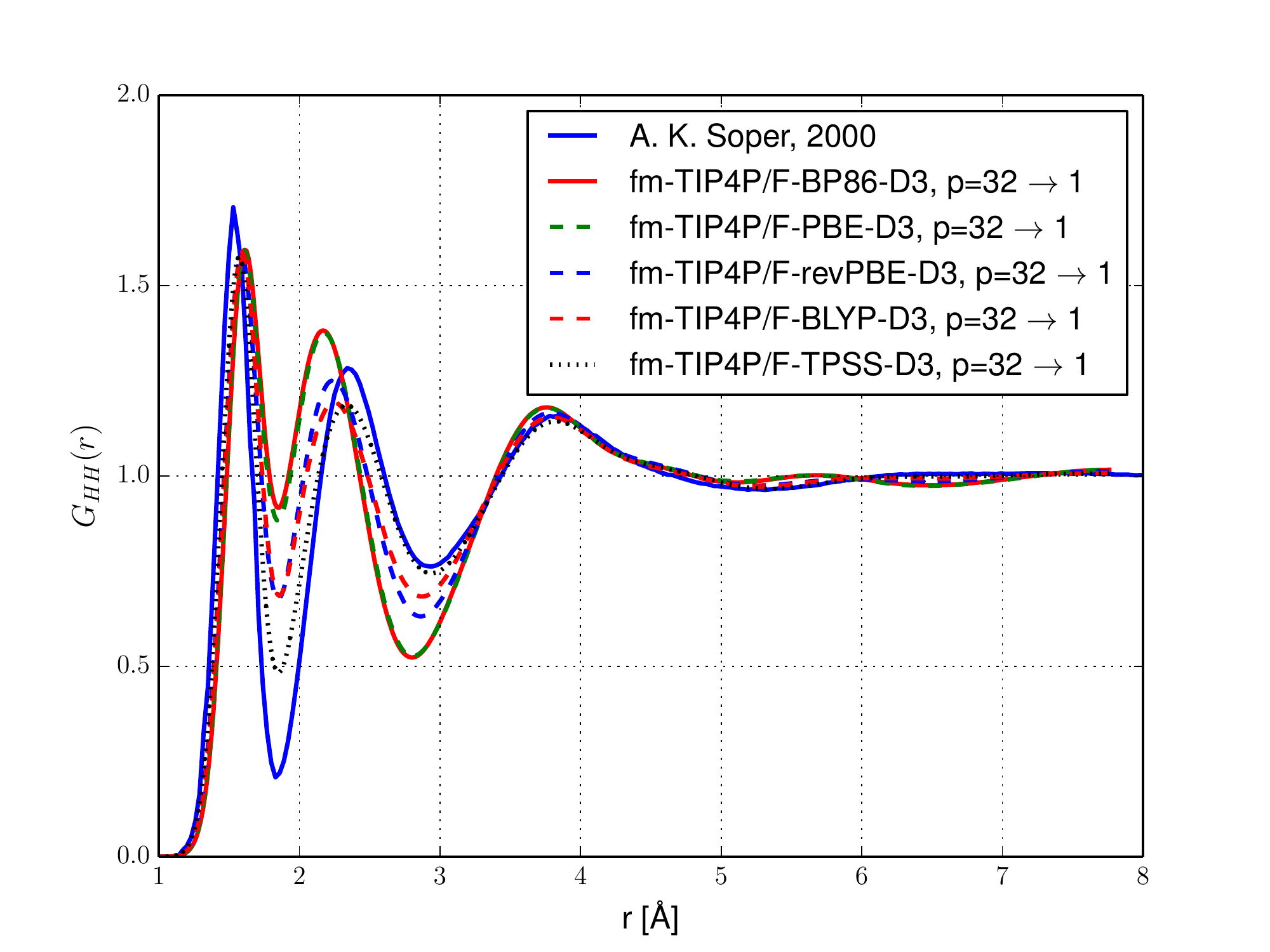}
    \caption{Hydrogen-Hydrogen RDFs from PIMD simulations using the fm-TIP4P/F-XC-D3 water model for the BP86, BLYP, revPBE, PBE and TPSS XC functionals, respectively. The experimental RDFs from Ref.~\onlinecite{soper2000radial} is shown for comparison\label{fig:comp_diff_func_HH}.}
\end{figure}

The force-matched water models shown in Table~\ref{tab:parameters} now allow us to investigate the influence of the various approximations to the XC functional, as reported in Fig.~\ref{fig:comp_diff_func_OO}, \ref{fig:comp_diff_func_OH} and \ref{fig:comp_diff_func_HH}, respectively. Taken together, these simulation results show that the RDFs calculated using the BP86-D3 XC functional is remarkably close to the the ones of the PBE-D3 functional, while the revPBE-D3, BLYP-D3 and in particular the TPSS-D3 XC functionals produced RDFs in increasing agreement with experiment. The former reflects the fact that the parameters of the fm-TIP4P/PBE-D3 and fm-TIP4P/BP86-D3 water potentials were rather similar to each other, as detailed in Table~\ref{tab:parameters}. All XC functionals led to water models with over-structured RDFs, as noted previously. Nevertheless, given that the present water models were all derived from semi-local DFT calculations, the fm-TIP4P/TPSS-D3 water model was altogether in remarkably good agreement with the experimental measurements. In fact, it turned out to be in much better agreement than a previous calculation using the TPSS XC functional, though without van der Waals correction and NQE, suggested \cite{JoostWater2005}. It not only qualitatively reproduced the various average bond lengths and the correct relative heights of the first two intermolecular peaks of the O-H RDF, but also the correct delocalization of the average intramolecular O-H bond length, as well as the second solvation shell of the O-O RDF. As a consequence, in spite of the observed variations, and given the challenge of simulating liquid water, the performance of semi-local DFT that is underlying the present water models, can be judged to be reasonably good.

%
%
\section{Results and Discussion\label{sec:results}}

The results so far have focussed on assessing whether the force-matching procedure produces reasonable water models, as well as the impact of nuclear quantum effects, van der Waals interactions and XC functional; these results have primarily focussed on the reproduction of the experimental partial RDFs for liquid water, which are often poorly reproduced by DFT-based AIMD simulations. In this section, we perform more extensive simulations of static and dynamic equilibrium properties for a range of force-matched water models that otherwise would have not been feasible by direct AIMD simulations; as noted above, the force-matched models considered here were all derived from DFT calculations which employed the ``D3'' London dispersion correction. 

\subsection{Static Properties}

%
\begin{table*}
\caption{Static equilibrium properties of the force-matched water models for the different semi-local XC functionals obtained from PIMD simulations in the constant-NPT ensemble: $p$ denotes the number of ring-polymer beads (or imaginary time slices), $r_{\text{OH}}$ the intramolecular O-H bond length, $\theta_{\text{HOH}}$ the H-O-H bond angle, $\mathcal{\mu}$ the molecular dipole moment, $\rho$ the equilibrium density and $\epsilon_{s}$ the static dielectric constant.} 
\label{tab:properties}
\begin{tabular}{l|c|c|c|c|c|l}
    XC Functional & $p$
    &$r_{\text{OH}}$\;[\AA]&$\theta_{\text{HOH}}$\;[deg]&$\mathcal{\mu}$\;[D]&$\rho$\;[g/cm$^3$]&$\epsilon_{s}$\\
    \hline\hline
    PBE-D3 \cite{pbe} & 1&0.9931&106.5224&2.1177&1.059&37.00\\
    PBE-D3 & 32$\to$1&1.0100&106.5183&2.1537&1.067&27.31\\
    BP86-D3 \cite{becke, perdew86}        &       1&0.9949&106.6028&2.1164&1.063&43.55\\
    BP86-D3  & 32$\to$1&1.0118&106.5948&2.1525&1.071&35.08\\
    BLYP-D3  \cite{becke, lyp}       &       1&0.9888&106.3276&2.0127&1.025&35.46\\
    BLYP-D3  & 32$\to$1&1.0048&106.3005&2.0460&1.030&31.35\\
    revPBE-D3 \cite{revPBE}      &       1&0.9863&106.0220&2.1012&1.011&40.05\\
    revPBE-D3       &32$\to$1&1.0042&106.0142&2.1396&1.018&35.77\\
    TPSS-D3   \cite{tpss}       &       1&0.9858&105.1119&2.1660&1.000&48.38\\
    TPSS-D3         &32$\to$1&1.0018&105.0494&2.2026&1.005&45.69\\
\hline
    q-TIP4P/F \cite{comp_quant_eff}
                    & 32$\to$1 & 0.978(1)      &    104.7(1)  &  2.348(1)      &   0.998(2)    & 60(3) \\
\hline
    Expt.  & $\cdots$     &0.97 \cite{soper2000radial}  &105.1 \cite{soper2000radial}  & 2.9(6) \cite{BadyalSoper2000} & 0.997 \cite{SaulWagner1989}&78.4 \cite{Fernandez1995} \\
\end{tabular}
\end{table*}

Molecular static equilibrium properties such as the intramolecular O-H bond length $r_{\text{OH}}$ and the H-O-H bond angle $\theta_{\text{HOH}}$, as well as the molecular dipole moment $\mathcal{\mu}$ are shown in Table~\ref{tab:properties}. We find that the inclusion of NQE increases $r_{\text{OH}}$, which is indeed in agreement with path-integral calculations of others \cite{lobaugh1997quantumwater, SternBerne2001, FanourgakisXantheas2008, PhysRevLett.91.215503, PhysRevLett.101.017801}, but our calculated values are larger than the experimental value.\cite{soper2000radial} However, NQE reduced $\theta_{\text{HOH}}$ in contrast with previous path-integral simulations,\cite{SternBerne2001, lobaugh1997quantumwater} but consistent with Ref.~\onlinecite{FanourgakisXantheas2008} and, more importantly, systematically improved the agreement with experiment \cite{soper2000radial}. We find that density also increases when NQE are included, which is again just like the flexible and polarizable TTM3-F water model of Fanourgakis and Xantheas \cite{FanourgakisXantheas2008}, though at variance with Paesani \textit{et al.} \cite{paesani:184507, paesani2007quantumeffects}. 
In addition, $\mathcal{\mu}$ is also enhanced upon inclusion of NQE, though it still substantially underestimated relative to the experimental value.\cite{BadyalSoper2000} While this is consistent with previous classical and DFT-based PIMD simulations \cite{SternBerne2001, PhysRevLett.91.215503, PhysRevLett.101.017801}, it is in contrast with CMD simulations of Voth and coworkers using empirical force-fields.\cite{lobaugh1997quantumwater, paesani2007quantumeffects} The fact that the dipole moment magnitude is smaller than the values of previous classical MD calculations using polarizable force-fields (2.5-2.85~D) \cite{SprikKlein1988, SprikWater1991, RickBerne1994, lobaugh1997quantumwater, DangChang1997, SternBerneFriesner2001, AMOEBA2003, YuVanGunsterenWater2004, FanourgakisXantheas2006, LamoureuxWater2006, paesani2007quantumeffects, KolafaWater2011} and semi-classical AIMD simulations (2.7-3.1~D) \cite{laasonen1993ailiqwater, PhysRevLett.82.3308, silvestrelli:3572, PhysRevLett.91.215503, PhysRevLett.98.247401} can thus be attributed to the lack of polarizability of the present fixed point-charge water model. 

%
%
\subsubsection{Dielectric Constant}

As well as a large permanent dipole moment, liquid water also displays a large static dielectric constant of $\epsilon_{s} = 78.4$.\cite{Fernandez1995} In fact, this is the highest of all polar solvents with comparable dipole moments, and can be associated with the presence of a macroscopically extended HB network.\cite{Kauzmann1969} However, calculating $\epsilon_{s}$ using 
\begin{equation}
\epsilon_{s} = \epsilon_{\infty}+\frac{4\pi\beta}{3 V}(\langle
\bm{\mu}_p\cdot\bm{\mu}_p\rangle-\langle
\bm{\mu}_p\rangle\cdot\langle\bm{\mu}_p\rangle)\, ,
\end{equation}
where $\epsilon_{\infty}$ is the infinite-frequency dielectric constant and $\bm{\mu}_p$ the total dipole-moment averaged over all imaginary-time slices $p$, requires a PIMD trajectory of several nanoseconds in length to converge.\cite{DeLeeuw1980, adams1981, neumann1983} Because it is not feasible to converge this property with DFT-based AIMD simulations,\cite{PhysRevB.56.12847, MarzariVanderbiltParrinello} only rather crude estimates ($\epsilon_{s}$ = 67-86) using Kirkwood's theory \cite{Kirkwood1939} are available from first principles calculations. \cite{PhysRevLett.98.247401, silvestrelli:3572}

In order to obtain full dielectric relaxation, we equilibrated the system for $2~\text{ns}$ before calculating $\epsilon_{s}$ as an ensemble average over an additional $5~\text{ns}$. The corresponding results for the various XC functionals we have considered are shown in Table~\ref{tab:properties}. The fm-TIP4P/F-TPSS-D3 water model, which was consistently in best agreement with experiment within the present force-matched water potentials, also exhibits the highest dielectric constant. However, it severely underestimates the experimental value, as well as those obtained with several other empirical force-fields.\cite{Guillot2002, VegaAbascal2011} We note that the higher dipole moment of polarizable water models typically results in a dielectric constant that exceeds experiment, with typical values being in the range $\epsilon_{s} = 79-116$. \cite{LamoureuxWater2006, AMOEBA2003, SprikWater1991, YuVanGunsterenWater2004, SternBerneFriesner2001, RickBerne1994, KolafaWater2011} This suggests that the central reason for the underestimation of the dielectric constant in the force-matched models is due to the relatively low molecular dipole moments, which are typically around 0.7~D lower than the experimental estimate. \cite{BadyalSoper2000} With this large difference in dipole moment, as well as clear differences in the liquid structure for these different models, it is not surprising that the DFT-based models developed here underestimate the dielectric constant. 

Interestingly, we found that NQE reduced $\epsilon_{s}$ even further, which is rather surprising since at the same time they enhanced $\mathcal{\mu}$, as well as $r_{\text{OH}}$ and thus the root-mean square total dipole moment. Due to the fact that the latter is proportional to $\epsilon_{s}$, this immediately suggests that NQE should have lead to a larger instead of a lower value. Nevertheless, this is consistent with previous CMD calculations using the SPC/F ($\epsilon_{s}$ = 94 $\rightarrow$ 74) and  SPC/Fw ($\epsilon_{s}$ = 80 $\rightarrow$ 64) water models \cite{lobaugh1997quantumwater, paesani:184507}, whereas the flexible and polarizable TTM2.1-F water potential of Fanourgakis and Xantheas \cite{FanourgakisXantheas2006} predicts a NQE induced increase of $\epsilon_{s}$ from 67 to 74.\cite{paesani2007quantumeffects}

%
%
\subsubsection{Density Maximum and Temperature of Maximum Density}

Due to the fact that the remaining calculations were computationally rather costly, we have restricted ourselves to simulations based on the fm-TIP4P/F-TPSS-D3 water potential, which has so far been found to give the overall best agreement with experimental properties, as noted above.

\begin{figure}
    \centering
    \includegraphics[width=0.5\textwidth]{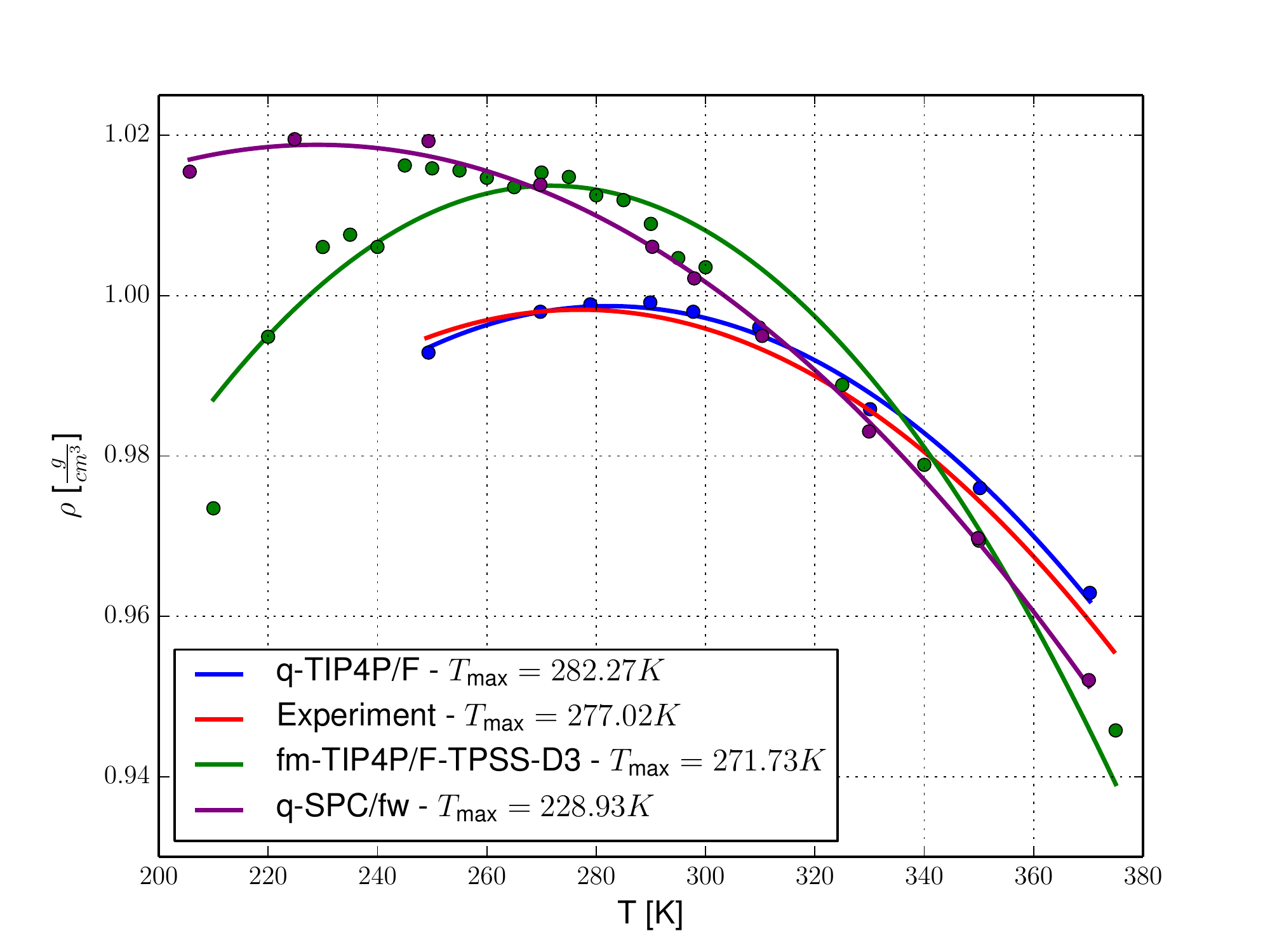}
    \caption{Liquid water density as a function of temperature for fm-TIP4P/TPSS-D3 water potential. The corresponding results of the q-TIP4P/F and q-SPC/F water models \cite{paesani:184507,comp_quant_eff}, as well as the experimental data,\cite{SaulWagner1989} are shown for comparison. 
    \label{fig:density}
    }
\end{figure}

To accurately calculate the average liquid density, we extended the equilibration time to $5~\text{ns}$ for temperatures below 280~K to account for the reduced molecular translational diffusion of undercooled water. The corresponding data points were fit to a parabola of the form $f(T) = a \left( T-T_{\text{max}} \right)^2 + \rho_0$ and are shown together with results from the q-SPC/Fw and q-TIP4P/F water models in Fig.~\ref{fig:density}.\cite{paesani:184507, comp_quant_eff} We find that the q-SPC/Fw force-field underestimates the experimental temperature of maximum density at $T_{\text{max}} = 277~K$ by as much as $\sim 48~K$, while results for q-TIP4P/F and the present fm-TIP4P/F-TPSS-D3 are in much better agreement with the experimental $T_{\text{max}}$. \cite{DonchevPNAS2006,paesani2007quantumeffects} The maximum density of the q-SPC/Fw and fm-TIP4P/F-TPSS-D3 water potentials, however, are somewhat too high, while the q-TIP4P/F is in excellent agreement with experiment.\cite{comp_quant_eff} The fact that including the ``D3'' London dispersion correction had the tendency to overcorrect the otherwise too low density of liquid DFT water is consistent with recent DFT-based AIMD simulations \cite{kuehnewaterinterface2011, TuckermanWater2012}.

%

%
%
\subsubsection{Melting Point}

We have performed PIMD simulations at atmospheric pressure to determine the quantum melting point of the fm-TIP4P/F-TPSS-D3 water model by direct coexistence simulations of the  water/hexagonal ice interface. Because liquid water has a higher density than hexagonal ice, we have chosen to use the simulation box density as an order parameter the distinguish between formation of solid hexagonal ice and liquid water.

%
\begin{figure}
\includegraphics[width=0.5\textwidth]{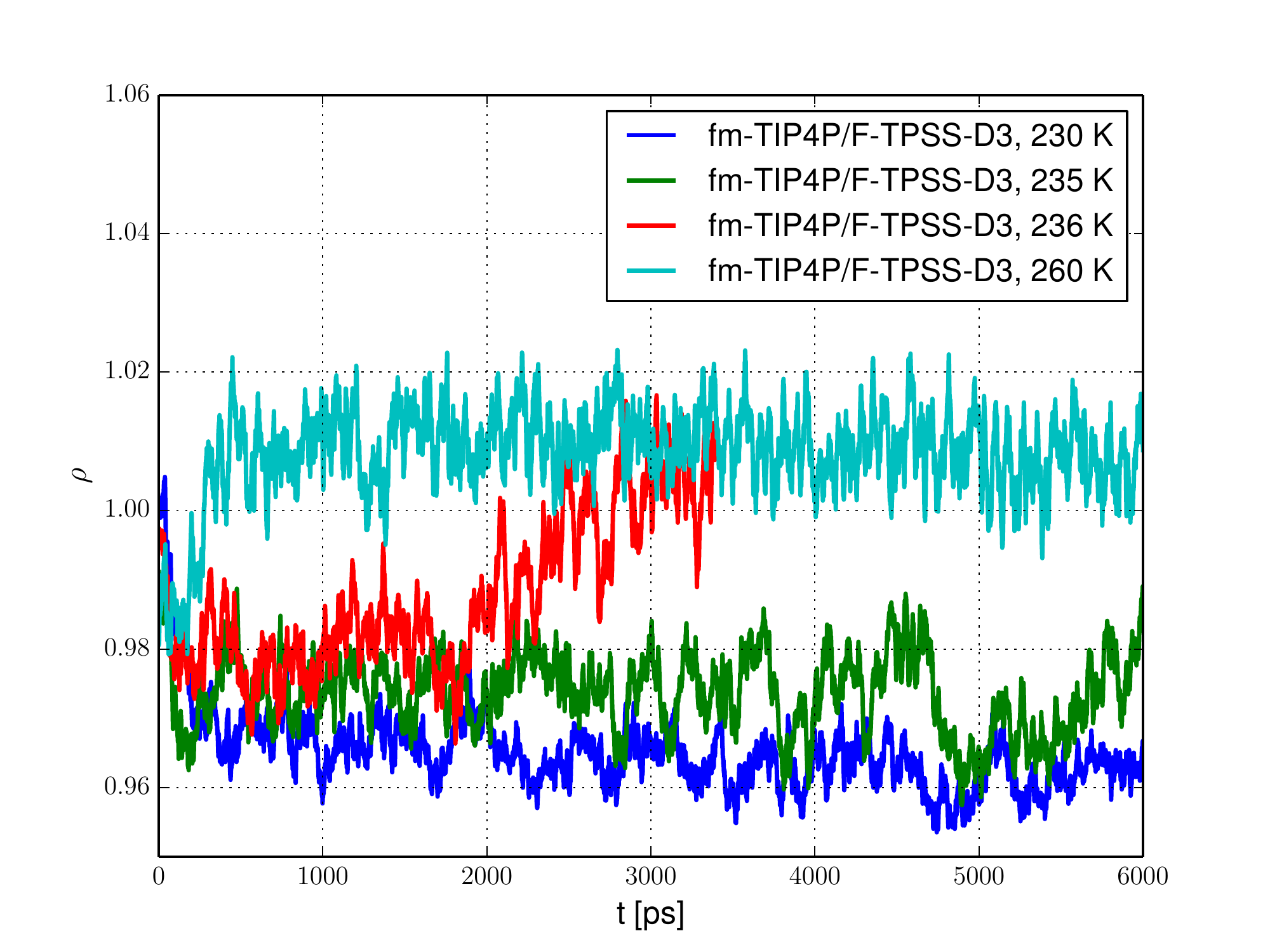}
\caption{Density profiles during PIMD simulations to determine the melting point of fm-TIP4P/F-TPSS-D3. At a temperature of 230~K, the system clearly remains in the ice phase. Just above 235~K, the ice phase melts and an higher (liquid) density is observed.}
\label{fig:melt_density}
\end{figure}

Figure~\ref{fig:melt_density} illustrates typical density traces as a function of time in these coexistence simulations. Below 235 K, we find that the system adopts a density of around 0.96~g~cm$^{-3}$, corresponding to that of hexagonal ice; however, a simulation run at 236K demonstrates that the system melts to form liquid water. As a result, the melting temperature of the fm-TIP4P/F-TPSS-D3 potential has been found to be between $235-236~\text{K}$, which is around 38~K lower than the experimental value. For comparison, classical MD simulations of common rigid water models have been found to give melting temperatures that range from about 146~K for TIP3P to 274~K for TIP4P/ice.\cite{vega2005melting, VegaAbascal2011} Including NQE by means of PIMD calculations resulted in a melting temperature of $251\pm1~\text{K}$ and $195\pm5~\text{K}$ for the q-TIP4P/F and q-SPF/Fw water potentials, respectively \cite{comp_quant_eff}. The corresponding values from DFT-based AIMD simulations, however, are much higher, namely 360~K with and 411~K without vdW correction \cite{YooXantheas2009, YooXantheas2011}. 

We note that previous classical MD simulations have suggested that it is not possible to reproduce the experimental difference of 4~K between the melting point of hexagonal ice and the temperature of maximum density using fixed point-charge models.\cite{VegaAbascal2005} In fact, the present PIMD simulations using the fm-TIP4P/F-TPSS-D3 water model predict a difference of 35~K between these two temperatures, which is within the 21-37 K range of typical differences found by classical MD simulations using empirical force-fields.\cite{vega2005melting} Since the average molecular dipole moment of ice is significantly larger than that of liquid water, indicating that significant charge redistribution occurs upon freezing \cite{Batista1998, Batista1999}, it indeed appears that an explicit treatment of electronic polarization will be needed to quantitatively reproduce the small temperature difference between the temperature of maximum density and the melting point of water. 

%
%
\subsection{Dynamic Properties}

\subsubsection{Translational Diffusion Constant}

For the calculation of the translational diffusion constant $D$, one should bear in mind that it is sensitive to finite-size effects which arise from the fact that a diffusing particle sets up a hydrodynamic flow which decays slowly as $r^{-1}$. In a periodically repeated system this leads to an interference between one particle and its periodic images. To account for this well-known finite-size-dependence, we have therefore performed two RPMD simulations of smaller systems (containing 216 and 343 water molecules) and extrapolated $D$ to the infinite system-size limit using the relation of D\"unweg and Kremer, \cite{dunweg1993polymermd,yeh2004systemsizediffusion}
\begin{equation}
    D_{\text{PBC}}(L) = D_\infty - \frac{kT\xi}{6\pi\eta L},
\end{equation}
where $D_\infty$ is the diffusion coefficient for an infinite system, $\eta$ is the translational shear viscosity, $L$ is the length of the periodic simulation box and $\xi = 2.837$ a numerical coefficient which depends on the geometry of the simulation cell.

We found $D_{\infty}^{qm} = 0.288$~\AA$^2$/ps for the fm-TIP4P/F-TPSS-D3 water model, which is 25~\% above the experimental value of 0.23~\AA$^{2}$/ps \cite{Price1999}. For comparison, the translational diffusion constant has been reported by others to be 0.19-0.548~\AA$^{2}$/ps using CMD and RPMD simulations, respectively.\cite{DonchevPNAS2006, lobaugh1997quantumwater, paesani:184507, pena:5992, paesani2007quantumeffects, comp_quant_eff} In any case, this demonstrates that our fm-TIP4P/F-TPSS-D3 model suggests that DFT water is indeed fluid (at least for this combination of XC functional and vdW corrections). \cite{kuehnewater2009, LeeTuckerman2007}

A further interesting result relates to the observed quantum effect, defined here as the ratio of the quantum and classical diffusion coefficients. In the original development of the q-TIP4P/F model, it was found that the quantum effect was around 1.15; this was much smaller than previous values of 1.38-1.58, which had been obtained for either rigid or harmonically-flexible fixed-charge water models \cite{paesani2007quantumeffects, paesani:184507, DonchevPNAS2006, pena:5992, lobaugh1997quantumwater}. The relatively small quantum effect for q-TIP4P/F was found to be due to the existence of a ``competition'' between intramolecular and intermolecular ZPE contributions; in particular, intermolecular hydrogen bonds are weakened by ZPE, leading to faster translational dynamics, but the strength of intermolecular interactions is increased by changes to the molecular dipole moment which arise due to the incorporation of intramolecular ZPE. In the present work $D_{\infty}^{cl} = 0.300$~\AA$^2$/ps, which is smaller than the value including NQE, such that $D_{\infty}^{qm}/D_{\infty}^{cl} = 0.96$. In other words, the fm-TIP4P/F-TPSS-D3 water model exhibits an ``inverse'' quantum effect, meaning that diffusion actually slows down when NQE are included. Although difficult to confirm without further detailed investigations, it seems that a likely explanation is the fact that the intramolecular potential in the force-matched fm-TIP4P/F-TPSS-D3 potential derived here is much ``softer'' than the original q-TIP4P/F model, as already noted above. As a result, the addition of intramolecular ZPE as one moves from classical simulation to one including NQE may have a larger impact on intermolecular forces than in the original q-TIP4P/F model; this effect, along with the overly-tetrahedral structure of the water model proposed here, may lead to the observation of an ``inverse'' quantum effect. Given the experimental evidence from isotopically-substituted water, where normal water diffuses faster than heavy water (D$_{2}$O), this suggests that there remain some feature of our empirical models which fails to account correctly for the influence of quantum fluctuations; investigating these features will be an aim of future work.

%
%
\subsubsection{IR Spectrum}

The IR absorption spectrum of liquid water at ambient conditions using the fm-TIP4P/F-TPSS-D3 water model was obtained as the Fourier transform of the dipole autocorrelation function 
\begin{equation}
n(\omega)\alpha(\omega)=\frac{\pi \beta \omega^2}{3 c V \epsilon_0} \tilde{c}_{\bm{\mu}\cdot\bm{\mu}}(\omega)\,.
\end{equation}
Here, the PACMD approximation to the Kubo-transformed dipole autocorrelation function $\tilde{c}_{\bm{\mu}\cdot\bm{\mu}}(t)$ was calculated using
\begin{eqnarray}
\mu_J(t)=\mu_J(\bm{R}_{J,O}^{(c)}(t),\bm{R}_{J,H_1}^{(c)}(t),\bm{R}_{J,H_2}^{(c)}(t)),
\end{eqnarray}
corresponds to the dipole moment operator of molecule $J$ evaluated at the centroid of the PACMD ring-polymer system at time $t$.

%
\begin{figure}
\includegraphics[keepaspectratio,width=1.15\linewidth,height=0.55\textheight]{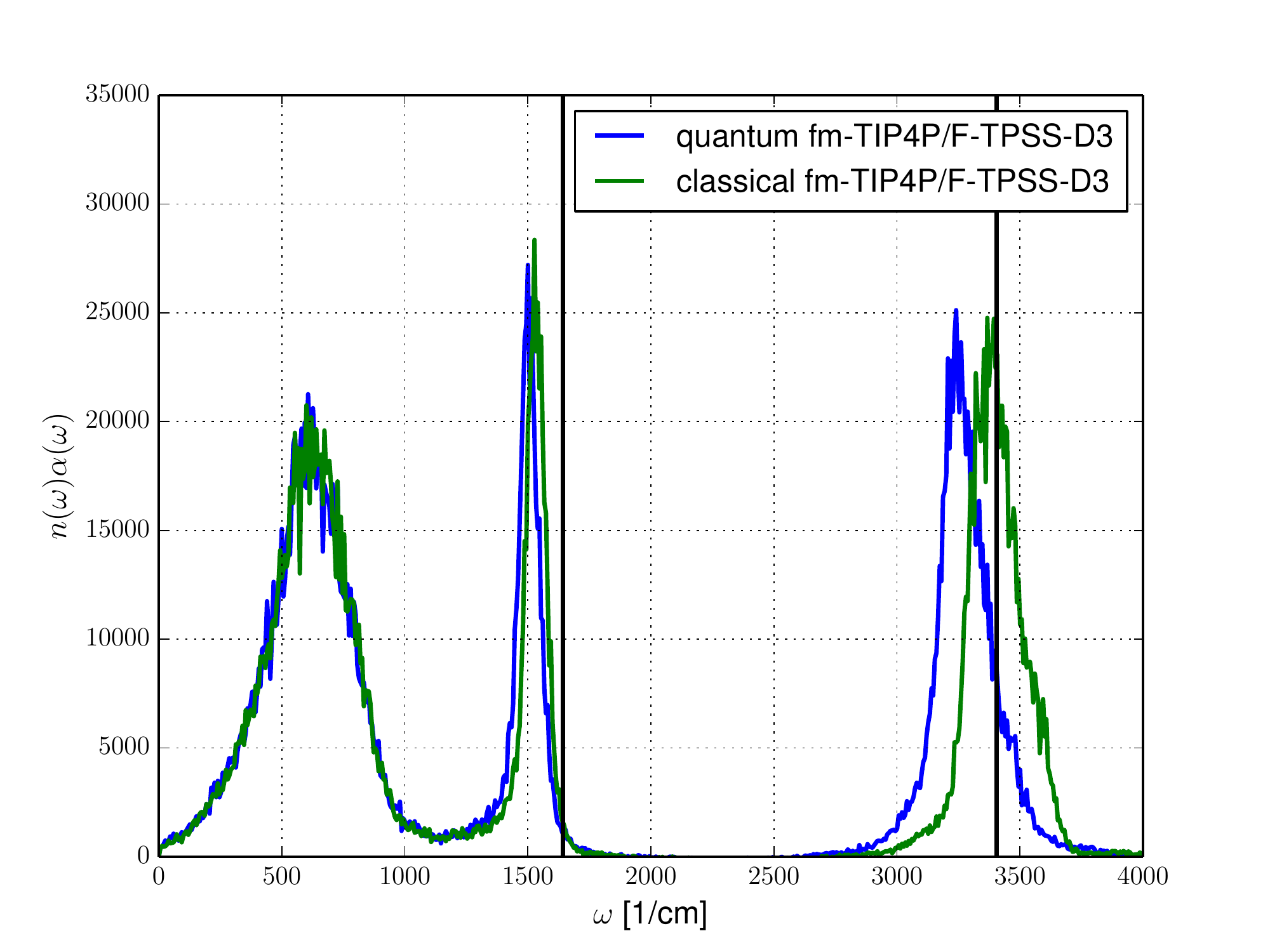}
\caption{Classical and quantum PACMD dipole absorption spectrum of the fm-TIP4P/F-TPSS-D3 water model. The experimental bulk water values from Ref.~\onlinecite{venyaminov1997absorptivity} are drawn vertically for comparison.}\label{fig:IR}
\end{figure}

The classical and quantum dipole absorption spectra of the fm-TIP4P/F-TPSS-D3 water model are compared in Fig.~\ref{fig:IR}. The two calculated IR spectra clearly reproduce the general features of the experimental spectrum, with O-H stretching absorptions above $\sim 3000~\text{cm}^{-1}$, a water bending band at around $\sim 1600~\text{cm}^{-1}$, and intermolecular librational features below $\sim 1000~\text{cm}^{-1}$. Moreover, the peak at $\sim 200~\text{cm}^{-1}$ is absent from both of the simulated spectra. This peak arises from the low-frequency modulation of dipole-induced dipole interactions which are clearly not present in simple point-charge models. \cite{ImpeyMadden1982}
However, the calculation including NQE shows the typical red-shift in comparison to the classical one, \cite{lobaugh1997quantumwater, paesani:184507, paesani2007quantumeffects, RosskyPNAS2005, habershonIR, FanourgakisXantheas2008} although we note that it remains unclear to what extent this is due to the well-known ``curvature problem'' observed by Marx and coworkers. \cite{WittRPMD2009, IvanovRPMD2010, WhMiller2009, WhMiller2011,PaesaniIR2010}
We find that, while the O-H stretching frequencies of the force-matched model reproduce the experimental values reasonably well, whereas the q-SPC/Fw water model predicts distinct antisymmetric and symmetric stretching peaks \cite{zhangwater2013}. However, the bending frequency is under-estimated by around $100~\text{cm}^{-1}$; again, this may be a simple consequence of the parameters determined in the force-matching procedure; all of the force-fields derived here exhibit bending force constants $k_{\theta}$ which are smaller than that of the original q-TIP4P/F model, which itself reproduces the experimental bending frequency quite accurately.

%
%
\section{Conclusion\label{sec:conclusion}}

In this paper, we have developed a series of q-TIP4P/F-like water models using a force-matching algorithm based on reference forces from DFT calculations. Subsequent classical and quantum simulations of the resulting water models demonstrated a wide range of results depending upon which exchange-correlation functional was employed in the calculation of the reference forces used as input for the force-matching procedure; however, some trends are apparent. Almost all force-matched water models resulted in over-structured liquid water when compared to experimental results; this finding is not uncommon in DFT-based simulations, so it is not surprising that empirical models based on DFT reference forces exhibit a similar propensity. 

Overall, we found that the fm-TIP4P/F-TPSS-D3 model offered the best agreement with experimental properties, including the density maximum, temperature of maximum density, melting point, translational diffusion constant and the IR spectrum. However, it is interesting to note that none of the force-matched models developed here offered performance on par with the original q-TIP4P/F water model; this may point to an insufficient accuracy of the DFT reference forces, but we must also bear in mind that the q-TIP4P/F force-field was designed to specifically reproduce experimental properties in quantum simulations, rather than being derived from \textit{ab initio} reference forces.

Despite this, there are many improvements which could be made to build on the present study. For example, the q-TIP4P/F-like models developed here clearly neglect polarisability, an effect which could be easily incorporated into the current force-matching scheme. Furthermore, the same fitting procedure could be applied to reference data obtained at a higher level of theory. Both of these are areas for future work.

%
%
\begin{acknowledgments}
We would like to thank Professor David Manolopoulos for many fruitful discussions regarding this paper. Financial support from the Graduate School of Excellence MAINZ, the IDEE project of the Carl Zeiss Foundation and the University of Warwick is kindly acknowledged. T.D.K. gratefully acknowledge the Gauss Center for Supercomputing (GCS) for providing computing time through the John von Neumann Institute for Computing (NIC) on the GCS share of the supercomputer JUQUEEN at the J\"ulich Supercomputing Centre (JSC).
\end{acknowledgments}

%
%

%

\end{document}